\documentclass{sig-alternate-05-2015}
\usepackage{amsmath}
\usepackage{times}
\usepackage{helvet}
\usepackage{courier}
\usepackage{paralist}
\usepackage{graphicx}
\usepackage{caption}
\usepackage{subcaption}
\usepackage{tabularx,booktabs}

\usepackage{url}
\usepackage{relsize}
\usepackage{hyperref}
\usepackage[svgnames]{xcolor}
\hypersetup{colorlinks=false, urlcolor=blue}

\usepackage{changepage}
\usepackage{tikz}
\usetikzlibrary{patterns, positioning, fit, calc}
\usepackage{helvet}
\usepackage[eulergreek]{sansmath}
\usepackage{xcolor,colortbl}
\usepackage{multirow}
\usepackage{ragged2e}
\usepackage{color,soul}
\usepackage{fixltx2e}
\usepackage{adjustbox}

\begin{document}

\CopyrightYear{2016}
\setcopyright{acmcopyright}
\conferenceinfo{HT '16,}{July 10-13, 2016, Halifax, NS, Canada}
\isbn{978-1-4503-4247-6/16/07}\acmPrice{\$15.00}
\doi{http://dx.doi.org/10.1145/2914586.2914588}

%\title{Is this a Suitable Place to Open my Caf{\'e}? \\Predicting a business' Success in an Urban City}
%\title{Predicting a Physical Store's Success through Location-Based Services Data from Facebook}
%\title{Predicting a Brick-and-Mortar Store's Success via Online Location-Based Services Data from Facebook}
%\title{Predicting a Brick-and-Mortar Store's \\Success Using Facebook Data}
%\title{Where Businesses Flourish: Predicting a \\Retail Store's Success Using Facebook Data}
%\title{How Popular Will My New Store Be? \\A Study Using Facebook Data}
%\title{How Popular Will My New Store Be? \\Predicting Success Using Facebook Data}
%\title{Spotting the Business Goldmine: \\Location Analytics Using Facebook Data}
%\title{Spotting the Goldmine: Business \\Location Analytics Using Facebook Data}
%\title{Unearthing the Gold Mine of Food Businesses: \\Business Location Analytics Using Facebook Data}
%\title{Where Should You Set Up Your Store? \\Business Location Analytics using Facebook Data}
\title{Where is the Goldmine? Finding Promising Business Locations through Facebook Data Analytics}

\author{
\alignauthor
Jovian Lin, Richard Oentaryo, Ee-Peng Lim, Casey Vu, Adrian Vu, Agus Kwee\\ %, Philips Prasetyo \\
%\affaddr{$^1$School of Computing, National University of Singapore, Singapore}\\
%\affaddr{$^2$Interactive and Digital Media Institute, National University of Singapore, Singapore}\\
\affaddr{Living Analytics Research Centre, Singapore Management University, 80 Stamford Road, Singapore}\\
\email{jovian.lin@gmail.com, \{roentaryo, eplim, caseyanhthu, adrianvu, aguskwee\}@smu.edu.sg}\\
}

\maketitle % make the title area

%==================%
% *** ABSTRACT *** %
%==================%
\begin{abstract}
If you were to open your own cafe, would you not want to effortlessly identify the most suitable location to set up your shop?
Choosing an optimal physical location is a critical decision for numerous businesses, as many factors contribute to the final choice of the location.
%These factors include traffic flow, location of competitors, location of anchor businesses (\textit{i.e.}, other businesses that attract customers that would be beneficial for one's own business), and zoning.
In this paper, we seek to address the issue by investigating the use of publicly available Facebook Pages data---which include user ``check-ins'', types of business, and business locations---to evaluate a user-selected physical location with respect to a type of business.
Using a dataset of 20,877 food businesses in Singapore, we conduct analysis of several key factors including business categories, locations, and neighboring businesses.
From these factors, we extract a set of relevant features and develop a robust predictive model to estimate the popularity of a business location.
Our experiments have shown that the popularity of neighboring business contributes the key features to perform accurate prediction. We finally illustrate the practical usage of our proposed approach via an interactive web application system.
\end{abstract}


%==========================%
% *** Various Sections *** %
%==========================%
%%%%%%%%%%%%%%%%
%              %
% INTRODUCTION %
%              %
%%%%%%%%%%%%%%%%
\section{Introduction}

%\richard{
%Comments from Richard:
%
%1) The novelty point is not strong enough. Need to send a stronger message on what's new with our work. We should highlight this in the abstract, intro, as well as related work.
%
%2) In comparison to the geo-spot paper (published in KDD 2013), we need to emphasize on some advantages of our work, such as: a) we can predict location popularity at a a more fine-grained level, b) some unique features that we use but they don't
%}

%===========================================================%
% Talk about brick-and-mortar stores and their main problem %
%===========================================================%
%Amid buzzwords that fuel retail-related discussions, it's easy to forget a vital aspect of retail success: Location.
%``Location'' is a stubborn truism of retail success
\textbf{Motivation}. Location is a crucial factor of retail success, as 94\% of retail sales are still transacted in physical stores \cite{Thau2015}. %\footnote{Forbes: \url{http://onforb.es/1k8VEQY}}
To increase the chance of success for their stores, business owners require not only the knowledge of \emph{where} their potential customers are,
but also their surrounding competitors and complementary businesses.
From the property owners' standpoint, it is also important to assess the potential success values of their property locations so as to determine the appropriate businesses to lease the locations to and for the right amounts. 
However, assessing and picking a store location is a cumbersome task for both business and property owners.  

To carry out the above tasks well, many factors need to be taken into account, each of which requires gathering and analyzing the relevant data.
Traditionally, business and property owners conduct surveys to assess the value of store locations \cite{Cohen2000}.  
Such surveys, however, are costly and do not scale up well.  
With fast changing environments (\textit{e.g.}, neighborhood rental, local population size, composition, etc.) 
and emergence of new business locations, one also needs to continuously reevaluate the value of store locations. %, which would incur more costs and efforts.

%===============================================%
% Mention how social networks/big data can help %
%===============================================%
Fortunately, in the era of social media and mobile apps, we have an abundance of online user-generated data, which capture both activities of users in social media as well as offline activities at physical locations.
Facebook is one of the world's largest social media platforms, with more than 1 billion active users everyday \cite{Smith2014}. 
%\footnote{Facebook User Statistics (May 2015): \url{http://bit.ly/1Dn7nYS}},
From the business standpoint, the massive availability of user, location, and other behavioral data in Facebook is attractive, and has changed the way people do businesses.
For instance, many small/medium business owners are now setting up Facebook Pages to:
(i) allow customers to find their businesses on Facebook;
(ii) connect with customers via ``likes'' and ``check-ins'';
(iii) reach out to more customers through advertising their business pages on Facebook; and
(iv) conduct analytics of their pages to get a deeper understanding of their customers and marketing activities.

Consumers are also adapting both their online and offline behaviors to the introduction of Facebook Pages for businesses.
Other than ``liking'' businesses on their Facebook Pages, they can do a ``check-in'' whenever they physically visit the respective
business stores. 
Facebook Pages have turned many offline signals into online behavior that can be analyzed for business insights.
In particular, features such as ``likes'' and ``check-ins'' can be used as indicators of popularity, and by extension, success.
Similarly, Instagram, Twitter, and Foursquare also have variants of these quantitative signals that can be retrieved from their
geotagged photos, tweets, and tips.
%Consequently, the location-based services in social media have accumulated massive volumes of data.
These data allow us to study the dynamics of brick-and-mortar stores and discover meaningful patterns and
insights that will help retail and property owners make better decisions.
%
%Fortunately, our world is flooded with data, from social media to satellite data,
%and the problem of assessing a location's suitability can be alleviated by
%discovering meaningful patterns in the data through spatial analysis,
%which allows us to quantify the implications, consequences, and impact
%of a given location's suitability, thereby optimizing the business owner's decisions.
%Furthermore, with the rising use of Facebook Pages in retail businesses,
%we now have more ubiquitous and easily accessible information of businesses.
%Facebook also turns many offline signals into online behavior that can be analyzed for business insights.
%For instance, features such as ``likes'' and ``check-ins'' can be used as indicators of a business' success, or popularity.
%From the standpoint of businesses, the proliferation of social networks into global,
%mainstream culture has changed the way people do businesses.
%Moreover, with the ubiquity of Facebook and today's increasingly social-media-savvy consumers,
%it is no longer a ``good idea'' for most businesses to be on Facebook ---
%with more than 890 million people actively using Facebook
%every day\footnote{\url{http://bit.ly/1Dn7nYS}},
%it has  become a go-to component of almost any inbound marketing strategy.

\textbf{Objective}. In this work, we make use of data collected from Facebook Pages to answer important research questions such as:
``\textit{Where should an owner set up his physical retail store at, so as to optimize the store's popularity?}'',
``\textit{What are the important factors influencing a store's popularity?}'',
and
``\textit{Is there a ``local'' effect, whereby businesses can benefit from the presence of more popular/established neighbors?}''
%For example, will a small family-owned cafe benefit from a nearby Starbucks?
%Will small food outlets benefit by situating themselves in sizable shopping malls with many other food outlets?
To this end, we propose a new location analytics framework that operates on top of Facebook Pages data. 
The centerpiece of our current framework is the following prediction task: \textit{Given a target location that a business/property owner wants to hypothetically set his/her store at, how can we extract the relevant data of businesses within the vicinity of the target location and use them to estimate the popularity of the target location?}

\begin{figure}[!t]
\centering
\setlength{\fboxsep}{0pt}
\setlength{\fboxrule}{0.5pt}
\fbox{\includegraphics[width=1.0\columnwidth]{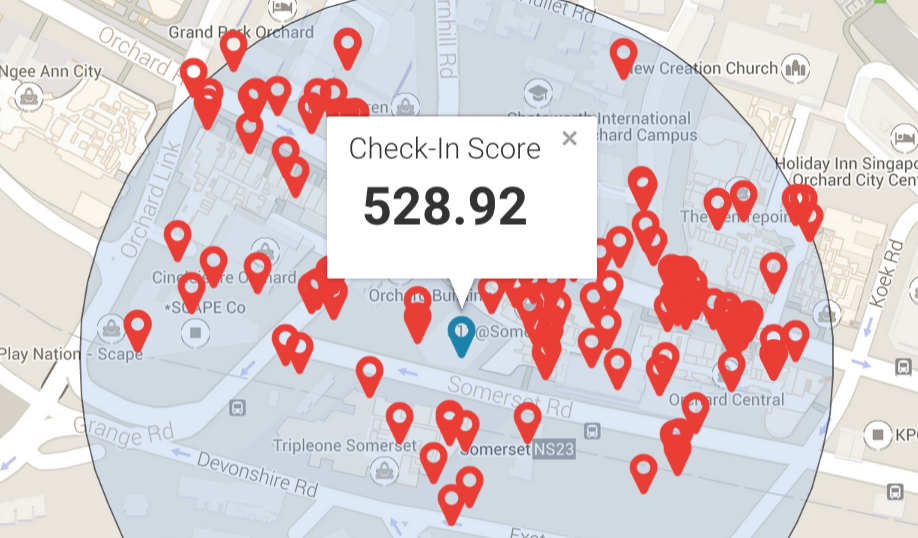}}
\caption{Prediction of ``check-ins'' based on the location indicated by the blue pin.
The red pins represent nearby existing businesses.}
\label{figure:framework_prototype}
\end{figure}

As an illustration, Figure~\ref{figure:framework_prototype} shows a visualization of our web application system that realizes our location analytics framework. 
For the system's input, a business/property owner first drops a blue pin on the map that indicates the hypothetical location of his/her new store. 
Our system then retrieves the relevant information about the area nearby, which are also occupied by the existing businesses as indicated by the red pins.
Based on these inputs, the system extracts a set of features and invokes a machine learning algorithm to predict the ``check-in'' score for the target location, which in turn serves as an indicator for the potential popularity of that location.
%The predicted ``check-ins'' of the given physical location is calculated based on the engineered features mined by incorporating characteristics of the area nearby, which is also occupied by several other existing stores indicated in red pins.

%++++++++++++++++%
% Figure (START) %
%++++++++++++++++%
%++++++++++++++%
% Figure (END) %
%++++++++++++++%

\textbf{Contributions}. In this paper, we show how publicly available Facebook Pages data can be analyzed and used to predict the potential popularity of a business location. 
To our best knowledge, this is the first work that demonstrates the feasibility of using Facebook data for business location analytics and, in particular, for aiding business and property owners to evaluate the value of a store location. 
It is also worth noting that %, in contrast to the existing works on location-based social networks \cite{Karamshuk2013,Georgiev2014}, 
our work presents a \emph{fine-grained} approach that allows the business/property owners to estimate the popularity of \emph{any} point on the city map. Our approach can be easily extended to predict multiple points simultaneously as well.
We summarize our main contributions as follows:

\begin{itemize}
\item We present a new study on business location analytics using Facebook Pages data. Specifically, we conduct detailed analyses on 20,877 Facebook Pages of food-related businesses in Singapore, which constitute one of the largest business types in the city with generally healthy visitor traffic. Based on the analyses, we identify key features that can be used to extract insights, as well as suitable metric for business popularity.

\item We develop a location analytics framework that includes a rich feature extraction module as well as a fast and accurate predictive model based on gradient boosting machine (GBM) \cite{Friedman:2001}. Unlike the previous close work on optimal store placement~\cite{Karamshuk2013} that outputs a ranked list of discretized areas (circles) with fixed radius, our approach is much more fine-grained. That is, our model can estimate on the fly the popularity of any arbitrary point on the map---which can be the location of an existing or a new/hypothetical business---without needing the locations to be discretized a priori.

\item Based on our (trained) predictive model, we analyze the contribution of key features that are crucial in predicting the popularity of a retail business---at both \textit{chunk} (\textit{i.e.}, a group of features) and \textit{individual} feature levels. In particular, we discover that distance-dependent features such as the total ``check-ins'' of businesses within certain radius are of utmost importance. We then provide an in-depth investigation on whether businesses, particularly smaller ones, benefit from the existence of other popular businesses within its  vicinity.

%\item We show a model that can predict a business' success based on publicly available information from Facebook.

%\item Through preliminary analysis, we discover that information such as the spatial distance between neighbors
%    provide powerful signals in the prediction task of popularity and success. We combine
%    the individual features using GBM to come up with a prediction model that can predict the ``check-ins''
%    of any business, given the latitude and longitude of a location.
%    %Experimental results show that GBM outperforms other regression methods. % baselines, such as nearest neighbors and Support Vector Regression.
    
% \item We discover that information such as the spatial distance between neighbors offer powerful signals to predict business popularity. 
%       We combine the individual features using GBM to come up with an \textit{accurate} predictive model that can estimate, at a fine-grained level, the 
%       ``check-ins'' of any business, given the latitude and longitude of a location. 
%       %Experimental results show that GBM outperforms other regression methods.

\item To concretely realize our idea, we have built an interactive web application that allows a business/property owner to drop a pin on the map and obtain a predicted ``check-in'' score for that location. The application is available at \url{http://research.larc.smu.edu.sg/bizanalytics/}.

% \item Our findings show how publicly available data from Facebook Pages can be exploited to predict popularity as well as the
% ``check-in'' score of a business --- existing or hypothetical.
% In that respect, retail owners could identify not only the good locations for setting up their business,
% but also compare how this predicted score compares with the surrounding neighbors such as competitors.

\end{itemize}

\textbf{Paper outline}. Section \ref{sec:related_work} first provides an overview of related works. In Section \ref{sec:dataset}, we describe the Singapore Facebook Pages data we use in our experiments. We subsequently elaborate our proposed location analytics approach in Section \ref{sec:framework}. Section \ref{sec:preliminaries} presents our experimental setup, followed by the results and analyses in Section \ref{sec:results}. We present our web application prototype in Section \ref{sec:prototype}, and finally conclude the paper in Section \ref{sec:conclusion}.

  % S1: Introduction
\section{Related Work}
\label{sec:related_work}

Our work can be viewed as a new type of \emph{location analytics} \cite{Garber2013}, which is an emerging area related to business intelligence (BI) \cite{Chen2012}. In recent years, organizations have relied on BI tools to delve into their data and reveal key insights that can aid their decision-making processes. With these tools, businesses have been able to make informed decisions based on what happened and when --- typically pertaining to sales figures and supplier transactions. 
%However, an equally important information that a growing number of organizations recognize is the question of \emph{where}. 
Lately, there is an important trend for organizations to address the question of \emph{where}.
Conventional BI systems, however, lack location-related analytics capabilities, 
and thus do not consider geographic and demographic factors  crucial for consumer analysis, \textit{e.g.}, where to set up stores, warehouses, or marketing campaigns.

Previously, combining separate BI and location-based approaches such as geographic information systems (GIS), was privileged only to large enterprises such as oil/gas-exploration companies, transportation companies, or government agencies \cite{Chen2012}. 
These technologies involve costly data acquisition processes and specialized labor skills. Moreover, their integration requires complex and time-consuming implementation. A recent survey by ESRI and IT Media firm TechTarget \cite{ESRI2012}
discovered that many organizations now believe that it is important to look at business data in a geographical context. Today, location-based data are abundant, thanks to the large volumes of user traces available from social media (such as Foursquare and Facebook) as well as mobile devices. 
However, many organizations are still unaware of the value of location-based data and struggle to put them to effective use. %, thus putting themselves at a competitive disadvantage.

%It has been established that geographical factors are highly correlated to the popularity of places \cite{Jensen2006,Porta2009}. A seminal study by Jensen \cite{Jensen2006} showed that %simple network analysis on pure location data suffice to unearth many important facts about the commercial organization of retail stores.
%Porta \emph{et al.} \cite{Porta2009} further showed that pure spatial organization can be indicative of the quality of retail and economic activities in several cities in Europe.
% Jovian: uncommented the above two citations

Using data from social media to understand the dynamics of a society
has always been a popular research theme,
particularly, in recommending a new location to a user.
For example, Facebook researchers Chang and Sun~\cite{location3:2011} analyzed Facebook users' ``check-ins'' data
to develop models that predict where users will ``check-in'' next.
They were able to predict user-to-user friendships (\textit{i.e.}, friend recommendation) just by the ``check-in'' data alone.
Gao \emph{et al.}~\cite{Gao:RecSys:2013} explored the use of Foursquare ``check-ins'' and temporal effects for
the task of location-recommendation; subsequently, the data were also used to predict
a user's location~\cite{Gao:CIKM:2013}.

Recent works on social media-based location analytics largely focus on detecting events and predicting user mobility patterns, although their use for BI applications are still limited so far. For instance, Li \emph{et al.}~\cite{Li2012} presented a machine learning method to discover and profile the user's location based on their following network and tweet contents. Noulas \emph{et al.}~\cite{Noulas2012} used Foursquare data to study the problem of predicting the next venue a mobile user will visit, by exploiting transitions between types of venues, mobility flows between venues, and spatio-temporal patterns of user ``check-ins''.
Also based on Foursquare data, Karamshuk \emph{et al.}~\cite{Karamshuk2013} demonstrated the power of geographic (\textit{e.g.}, types and density of nearby places) and user mobility (\textit{e.g.}, transitions between venues or incoming flow of users) features in predicting the best placement of retail stores. 
In a similar vein, Georgiev \emph{et al.}~\cite{Georgiev2014} conducted a study to predict the rise and decline of popularity of the local retail shops during the 2012 London Olympic Games. 
%Specifically, using a combination of geospatial and mobility features, they built a binary classifier to predict whether or not the number of customers in a venue during the Olympic Games will be higher than the expected number prior to the event.
Most recently, Zhang \emph{et al.} \cite{Zhang2016} extracted traffic and human mobility features from Manhattan restaurants data and studied how static and dynamic factors affect the economic outcome of local businesses in the city.

\textbf{Our approach}. Our work differs from all the above studies in several ways. Firstly, to our best knowledge, our work is the first to explore the use of Facebook data in business-location analytics. 
%To date, Facebook is the largest social graph in the world \cite{Adweek2014}, and has richer data than other location-based social networks such as Foursquare. %\footnote{\url{http://www.adweek.com/socialtimes/largest-social-networks-worldwide}}.
%With over 1.55 billion users visiting Facebook every month, of which 65\% check Facebook everyday \cite{FBBusiness2015}, %\footnote{\url{https://www.facebook.com/business/products/pages/}},
% https://www.quora.com/Are-Foursquare-and-Gowalla-going-to-survive-now-that-Facebook-Places-has-launched
With 1.55 billion monthly active users and 50 million business pages \cite{Smith2014}, Facebook can provide a more comprehensieve database of crowdsourced locations than other platforms (by comparison, Foursquare only has 55 million monthly active users and 1.3 million business pages \cite{Smith2014b}).
Secondly, instead of recommending places for users to establish retail stores or analyze on how unique events will affect businesses,we predict the popularity score of a user-selected venue, giving the user more freedom to choose \emph{anywhere} he/she wants to set up his/her business.
Thirdly, among all the works, Karamshuk \emph{et al.}'s~\cite{Karamshuk2013} is the closest to ours.
But the key difference is that their work \emph{discretized} the city into multiple circles with fixed radius and treated the issue as a ``ranking problem'', \emph{i.e.}, producing a ranked list of discretized circles.
In contrast, we view it as a ``prediction problem'' and provides a much more \textit{fine-grained}
approach of estimating the popularity of \textit{any} point on the map. Our method also works robustly on a range of radius values, instead of relying on a single predefined radius as in \cite{Karamshuk2013}.  
\section{Facebook Pages Dataset}
\label{sec:dataset}

In this section, we  first provide an overview of the data that we collected from Facebook, and then describe the important attributes found in the data. 
We then conduct a simple analysis on the two popularity measures---``check-ins'' and ``likes''---to determine the better metric for quantifying the popularity of a business.

\subsection{Data Harvesting}
%Facebook was our primary choice, among other location-based services such as Foursquare,
%as it is the largest social graph in the world\footnote{\url{http://www.adweek.com/socialtimes/largest-social-networks-worldwide}}.
%With over 1.35 billion people visiting Facebook every month,
%whereby 64\% check Facebook every day\footnote{\url{https://www.facebook.com/business/products/pages/}}, 
%% https://www.quora.com/Are-Foursquare-and-Gowalla-going-to-survive-now-that-Facebook-Places-has-launched
%Facebook can provide an extremely complete and continually-updated database of 
%crowdsourced locations.
In this paper, we focus our studies on food-related businesses found in the Facebook of Singapore. We choose food because it constitutes one of the largest business types in Singapore with generally healthy visitor traffic (``check-ins'' and ``likes'').
The food-related businesses were defined based on a manually-curated list that consists of $133$ food-related categories of business, such as those containing the words ``restaurant'', ``pub'', ``bar'', etc. 
%\richard{[Jovian: Please check if this is correct; I got this list from Casey]}.
%Jovian: yes it's ok.
In particular, we consider food-related businesses in Singapore Facebook Pages that explicitly specify latitute-longitude coordinates, and these coordinates must be within the physical boundaries of Singapore.
Using Facebook's Graph API \cite{FBGraph2015}, %\footnote{\url{https://developers.facebook.com/docs/graph-api/reference/page}},
we obtained a total of 82,566 business profiles within Singapore boundaries, 
of which we categorically filtered 20,877 (25.2\%) profiles 
that are food-related. %\footnote{We also, between Facebook and Foursquare, we were able to collect 43.9\% more food-related businesses in Singapore with Facebook (20,877 profiles) than with Foursquare (14,510 profiles).}.
All business data were analyzed in aggregate, and no personally-identifiable information was used.

\begin{figure}[!t]
\centering
\setlength{\fboxsep}{0pt}
\setlength{\fboxrule}{0.5pt}
\fbox{\includegraphics[width=1.0\columnwidth]{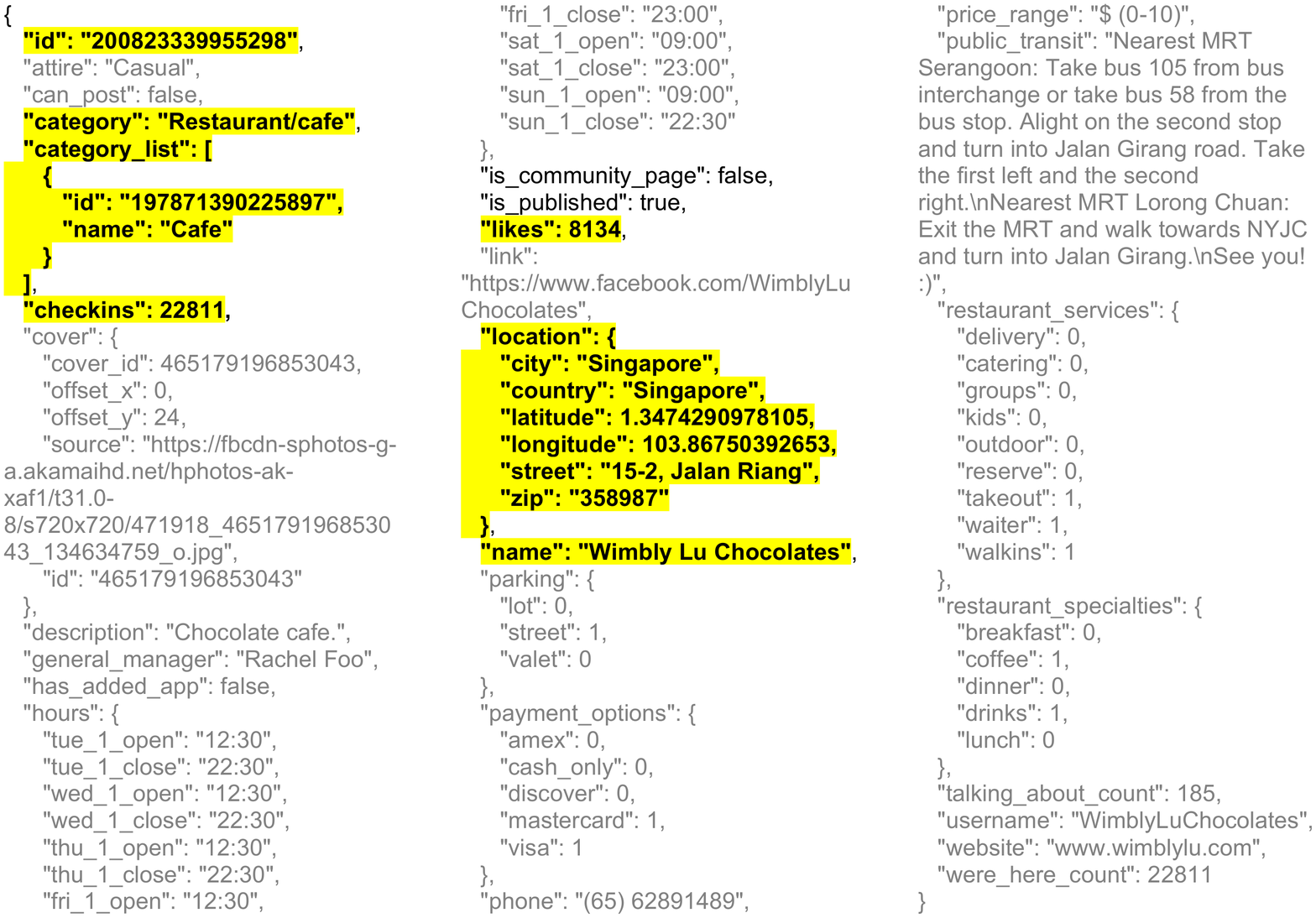}}
\caption{Facebook Graph API provides a JSON-formatted data of a business.
Shown here is the Facebook Page of the \textit{Wimbly~Lu Chocolates} cafe in Singapore.
%Source: \href{https://graph.facebook.com/200823339955298/}{\texttt{https://graph.facebook.com/200823339955298/}}}
Source: \url{https://graph.facebook.com/200823339955298/}}
\label{figure:wimbly_lu_json}
\vspace{2.5mm}
\end{figure}

\begin{figure}[!t]
\centering
\setlength{\fboxsep}{0pt}
\setlength{\fboxrule}{0.5pt}
\fbox{\includegraphics[width=0.63\columnwidth]{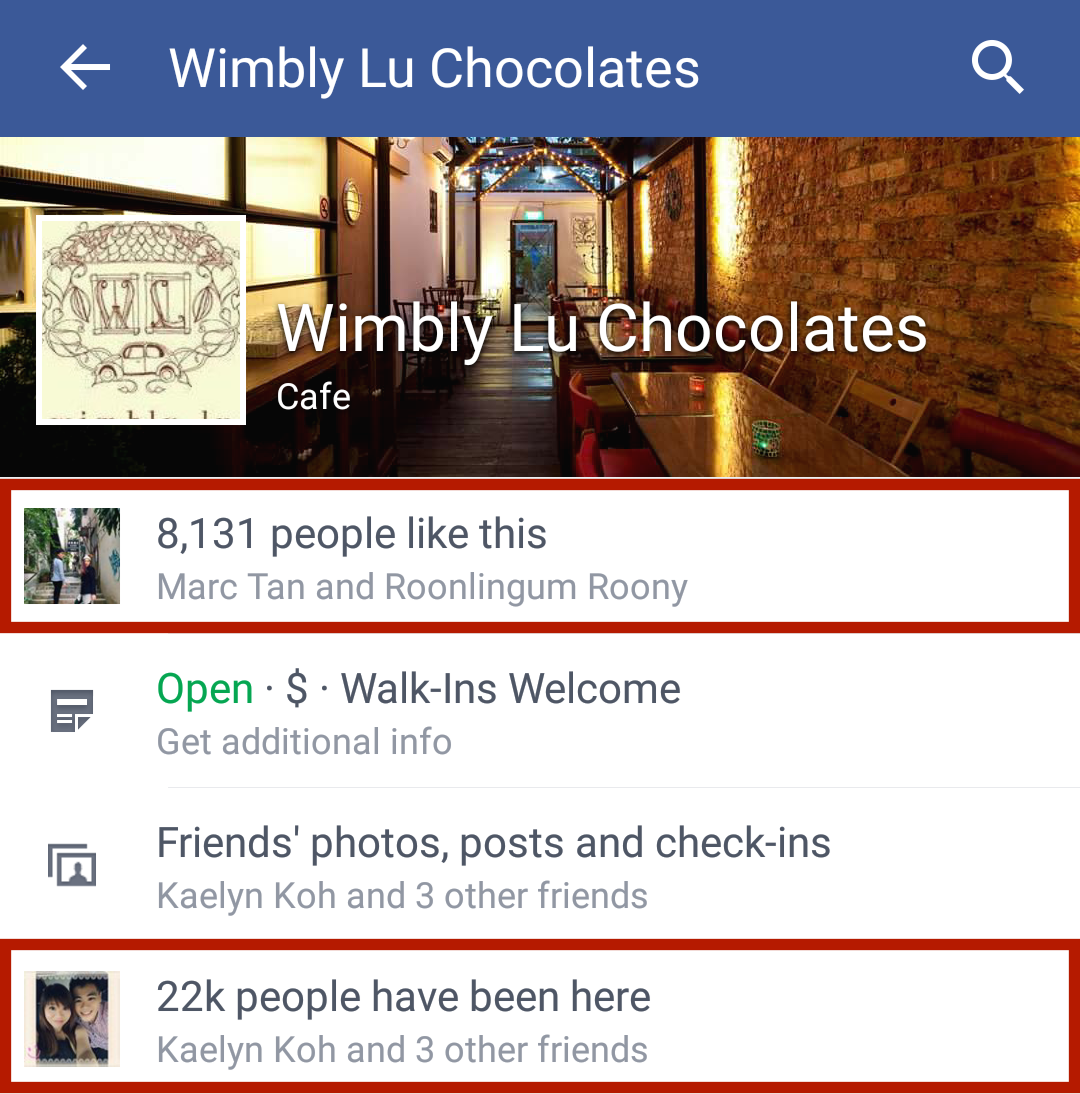}}
\caption{The Facebook page of \textit{Wimbly Lu Chocolates} that shows key popularity metrics such as 8.131 ``likes'' and 22K ``check-ins''.}
\label{figure:wimbly_lu_fb_page}
\end{figure}

Figure~\ref{figure:wimbly_lu_json} shows an example of one such 
business profile, \textit{Wimbly Lu Chocolates},
with important attributes (highlighted in bold) such as: 
\begin{inparaenum}[\upshape(\itshape i\upshape)]
\item ID,
\item category (\textit{i.e.}, the primary-category),
\item category list (\textit{i.e.}, the sub-categories), 
\item ``check-in'' count,
\item ``like'' count, and
\item location (including latitude and longitude).
\end{inparaenum} 
Figure~\ref{figure:wimbly_lu_fb_page} shows the corresponding Facebook Page of the business profile.

%+++++++++++++++++++++++++++++++++++%
% Figure (START) --- wimbly_lu_json %
%+++++++++++++++++++++++++++++++++++%
%++++++++++++++%
% Figure (END) %
%++++++++++++++%

%++++++++++++++++++++++++++++++++++++++++++++%
% Figure (START) --- wimbly_lu_fb_page_phone %
%++++++++++++++++++++++++++++++++++++++++++++%
%++++++++++++++%
% Figure (END) %
%++++++++++++++%

%++++++++++++++++%
% Figure (START) %
%++++++++++++++++%
% \begin{figure}[t]
% \centering
% \setlength{\fboxsep}{0pt}
% \setlength{\fboxrule}{0pt}
% \fbox{\includegraphics[width=0.48\textwidth]{figures/top_food_categories}}
% \caption{Top categories from food business profiles in Singapore.}
% \label{figure:top_food_categories}
% \end{figure}
%++++++++++++++%
% Figure (END) %
%++++++++++++++%

%+++++++++++++++%
% Table (START) %
%+++++++++++++++%
\begin{table*}[!t]
\centering
\small
\caption{Top 25 categories of food-businesses, their expected ``check-ins'', and those that perform better than the expectation.}
\label{table:business_categories}
\begin{tabular}{lcccc}
\toprule
\textbf{Categories} &  \textbf{Businesses} & \textbf{Total ``check-ins''} & \textbf{Expected ``check-ins''} & \textbf{\% of Businesses that have ``check-ins''} \\
                    &  \textbf{count}       &                          & \textbf{per business}       & \textbf{above the expected ``check-ins''} \\
\midrule
    Food \& Restaurant &             6,758 &           5,771,148 &              853.97 &                                             13.36\% \\
           Restaurant &              5,233 &           8,195,356 &             1,566.09 &                                             16.09\% \\
                 Cafe &              3,126 &           3,799,849 &             1,215.56 &                                             19.10\% \\
        Shopping Mall &              3,101 &           5,772,105 &             1,861.37 &                                             15.90\% \\
          Coffee Shop &              2,959 &           2,395,000 &              809.40 &                                             13.99\% \\
 Fast Food Restaurant &              2,840 &           3,447,999 &             1,214.08 &                                             19.12\% \\
       Food \& Grocery &             1,449 &           1,055,175 &              728.21 &                                              6.21\% \\
               Bakery &              1,338 &            394,982 &              295.20 &                                             11.58\% \\
   Chinese Restaurant &              1,157 &           2,376,432 &             2,053.96 &                                             23.16\% \\
           Food Stand &              1,099 &           1,454,820 &             1,323.77 &                                             10.92\% \\
                  Bar &               956 &           3,234,860 &             3,383.74 &                                             20.08\% \\
  Japanese Restaurant &               922 &           1,297,228 &             1,406.97 &                                             22.02\% \\
        Train Station &               879 &            429,740 &              488.90 &                                             26.17\% \\
            Nightlife &               744 &            866,619 &             1,164.81 &                                             10.62\% \\
        Movie Theater &               717 &           1,015,632 &             1,416.50 &                                              9.09\% \\
            Cafeteria &               661 &            421,774 &              638.08 &                                             11.20\% \\
   Seafood Restaurant &               629 &           1,786,933 &             2,840.91 &                                             22.58\% \\
   Italian Restaurant &               459 &            735,620 &             1,602.66 &                                             27.67\% \\
      Thai Restaurant &               437 &            593,546 &             1,358.23 &                                             26.32\% \\
     Ice Cream Parlor &               413 &            744,514 &             1,802.70 &                                             18.40\% \\
     Sushi Restaurant &               380 &            741,305 &             1,950.80 &                                             26.58\% \\
                  Pub &               369 &            513,009 &             1,390.27 &                                             20.33\% \\
           Night Club &               361 &           1,416,278 &             3,923.21 &                                             14.40\% \\
    Indian Restaurant &               350 &            538,624 &             1,538.93 &                                             18.00\% \\
\bottomrule
\end{tabular}
\end{table*}
%+++++++++++++++%
% Table (END) %
%+++++++++++++++%

\subsection{Categories Data}
\label{subsection:categories_data}

From the 20,877 food-related businesses, we retrieved a total of 357 unique categorical labels (as standardized by Facebook) 
from the attribute ``category~list'', which represents the sub-categories of a business. 
%We ignore the attribute ``category'' (note: lacking the additional ``list'') as this attribute more than 80/% of the businesses contains the label ``local business'',
%which is lacks comprehensiveness.
%Figure~\ref{figure:top_food_categories} shows the top 15 categories.
These categories contain not only food-related labels
(\textit{i.e.}, ``bakery'', ``bar'', ``cafe'', ``coffee shop''),
but also non-food labels such as ``movie theatre'', ``shopping mall'', and ``train station.''
The existence of non-food related labels within food businesses is
Facebook's way of 
allowing business owners to choose more than one categorical label for their business profile.
For example, a Starbucks outlet located at a train station in an airport would
likely have a mixture of food and non-food labels, 
such as ``airport'', ``cafe'', ``coffee shop'', and ``train station.''
% <para connect>
In addition, there is an intimate relationship between
the categories of a target business and those of its neighbors.
For instance, a family-run cafe will unlikely set itself next to an established coffee franchise like Starbucks,
whereas a dessert shop may be located near complementary dining places.

Table~\ref{table:business_categories} shows the top 25 categories of the food businesses in Singapore,
their expected ``check-ins'', and the percentage of businesses that perform better than expectation.
The proportion of businesses performing better than expectation ranges from 6 to 28\%.
The largest category is ``food and restaurant'', which is the most common category-type.
From the low percentages of those that actually perform better than the expectation,
we can tell that businesses obey a \emph{long-tail} distribution, with  the majority of businesses being unable to achieve the expected ``check-ins'' or more.

\subsection{Location Data}  %\subsection{Spatial Distance between a Business and its Neighbors}
\label{sec:location_data}

Each business profile has a \textit{location} attribute that contains
the physical address and latitude-longitude coordinates (hereafter known as ``lat-long'').
Knowing the location of every business allows us to calculate the neighborhood of a selected business
through the spatial distribution of other businesses around the vicinity.
Specifically, we consider the set $P_l = \{p | dist(p,l) \leq r \}$ of places $p$ that lie in a radius $r$ around a target location $l$.
The term $dist(p,l)$ denotes the \emph{Haversine distance}~\cite{Sinnott1984} between two locations $p$ and $l$. %and $P$ the set of food businesses in Singapore.
We can then create a two-dimensional \textit{distance matrix}
%(shown in Table~\ref{table:distance-matrix-between-businesses})
containing the distance between every pair of business.
For efficiency, we only consider a maximum radius of 1km (\textit{i.e.}, $r \leq 1$km).
This allows us to quickly retrieve the $k$ nearest neighbors of any location.

%+++++++++++++++%
% Table (START) %
%+++++++++++++++%
%\begin{table}[t]
%\scriptsize
%\caption{The distance matrix between businesses (or ``biz''). 
%Each element in the matrix represents the distance, in kilometers, between business A and business B.}
%\begin{tabular}{c|ccccc}
%~          & $biz_{1}$  &  $biz_{2}$  &  $biz_{3}$  &  \ldots  &  $biz_{N}$ \\
%\hline
%$biz_{1}$  & 0          & 0.612156    &  0.802870   &  \ldots  &  0.248020  \\
%$biz_{2}$  & 0.229087   & 0           &  0.991004   &  \ldots  &  0.399217  \\
%$biz_{3}$  & 0.193478   & 0.732814    &  0          &  \ldots  &  0.555121  \\
%\vdots     & \vdots     & \vdots      & \vdots      &  \vdots  &  \vdots    \\
%$biz_{N}$  & 0.556672   & 0.234799    &  0.902101   &  \ldots  &  0         \\
%\end{tabular}
%\label{table:distance-matrix-between-businesses}
%\end{table}
%+++++++++++++%
% Table (END) %
%+++++++++++++%

\subsection{Visitor Data}
Ideally, we would like to analyze customer information,
such as \emph{who} commented on or ``liked'' a business' Facebook post,
and match/recommend some user profiles to some businesses. % using, for instance, collaborative filtering~\cite{Koren:2011}.
However, due to privacy concerns, Facebook does not allow us to identify \emph{who} has checked-in or liked a particular business.
Although one can still crawl the posts on a business' wall, as of Facebook's Graph API v2.0 \cite{FBPlatform2015} (released in 2014), Facebook no longer supplies a user's actual ID.
Instead, Facebook uses the concept of ``app-scoped user IDs'', whereby a user's ID is unique to each app and cannot be used across different apps. %\footnote{\url{https://developers.facebook.com/docs/apps/upgrading\#upgrading_v2_0_user_ids}}.
As our crawler is considered an app, and Facebook limits the number of user posts that an app can query in a day, 
we are unable to gather enough posts---and by extension user IDs---to cover all (food-related) businesses in Singapore.
Having multiple crawlers will not work either, as the same user ID will be different for any two crawlers.

\subsection{Popularity Indicator: ``Check-in'' vs ``Like''}
%\subsection{Hotspot Data}

Facebook provides two possible indicators for a business page's popularity: ``check-in'' and ``like''. 
The ``check-in'' metric is common in location-based social media like Facebook and Foursquare.
Meanwhile, the ``like'' metric (shown as a ``thumbs up'' button) is more unique to Facebook, allowing users to express their recommendation/support for an entity.
%Both the number of ``check-ins'' and ``likes'' are signals that reflect the popularity  of a particular business and the online support that it has, respectively.
A ``check-in'' is the action of registering one's physical presence, and the total number of ``check-ins'' received by a business gives us a rough estimate of how popular and well-received it is.
In contrast, the number of ``likes'' literally reflects an online vote for the business.
Intuitively, therefore, a ``check-in'' should be a more suitable measure of a physical store's popularity, as it indicates a physical presence. 
%That is, the ``check-in'' count allows us to track how  \emph{many times} users physically visit a place.
Furthermore, ``check-in'' can be repeated, \emph{i.e.}, a user could ``check-in'' to a place on Monday and do so again on Tuesday.
By contrast, ``likes'' cannot be done repeatedly---it is a one-time event.

%Yet, for the sake of completeness, we investigate, among ``check-ins'' and ``likes'', which is the better  popularity feature.
%Pearson's correlation coefficient is a measure of how well your data would be fitted by a linear regression. If you only provide it with two points, then there is a line passing exactly through both points, hence your data perfectly fits a line, hence the correlation coefficient is exactly 1.

\begin{figure}[!t]
\centering
\begin{subfigure}{.23\textwidth}
 \centering
 \includegraphics[width=1\textwidth]{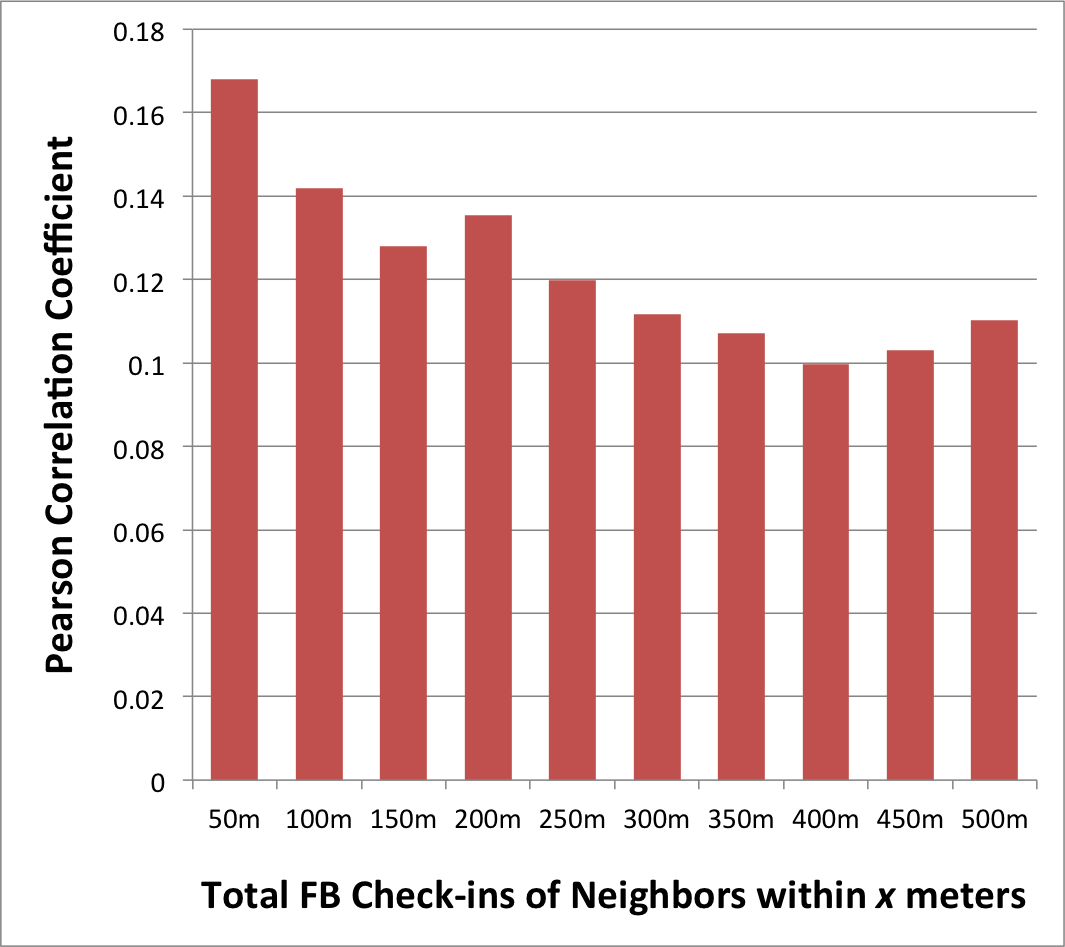}
 \caption{Target business' total ``check-ins'' vs. neighbors' total checkins}
 \label{fig:sub1}
\end{subfigure}
\begin{subfigure}{.23\textwidth}
 \centering
 \includegraphics[width=1\textwidth]{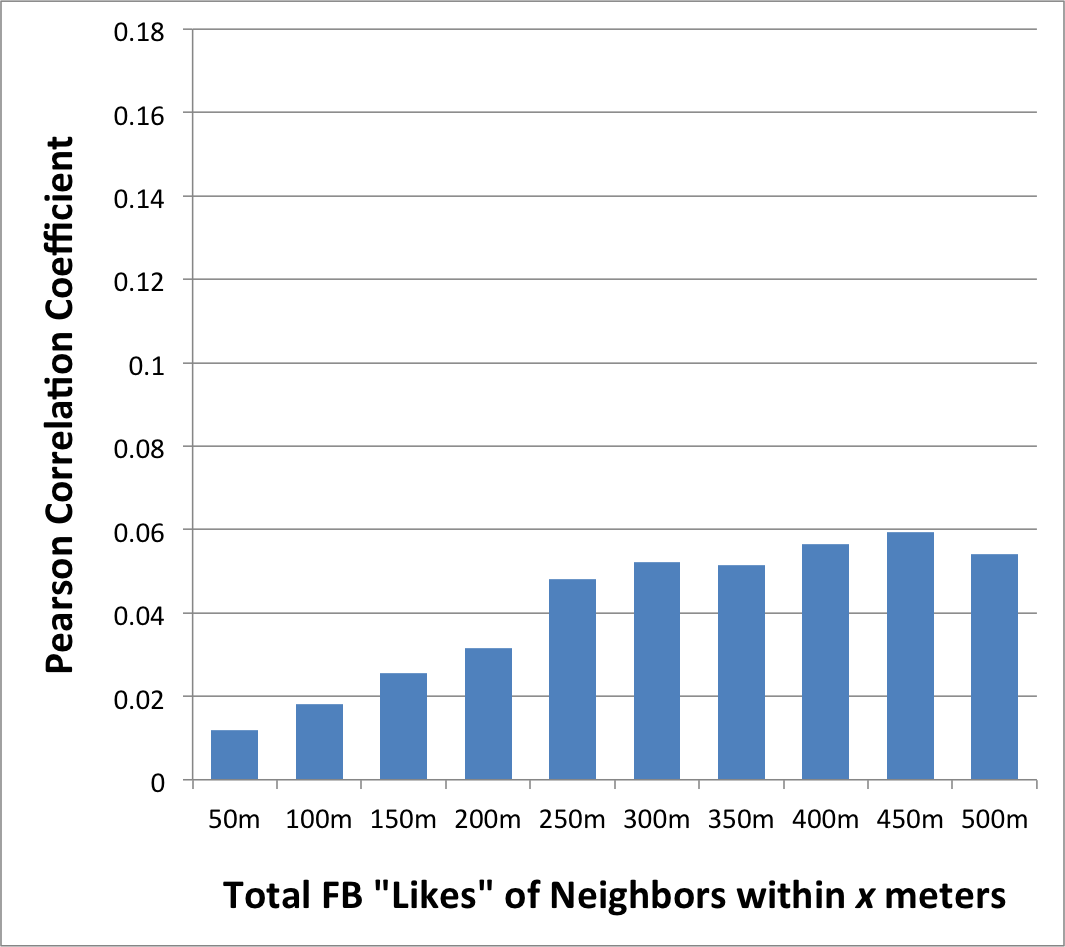}
 \caption{Target business' total ``check-ins'' vs. neighbors' total likes}
 \label{fig:sub2}
\end{subfigure}
\caption{Pearson correlation coefficient (PCC) of neighboring total ``check-ins'' (in red) and neighboring total ``likes'' (in blue).}
\label{figure:pcc}
\end{figure}

To prove this point, we compute the Pearson correlation coefficient (PCC) on two pairs:
\begin{inparaenum}[\upshape(\itshape i\upshape)]
\item the target business' ``check-ins'' w.r.t. its neighbors' total ``check-ins'', and
\item the target business' ``check-ins'' w.r.t. its neighbors' total ``likes''.
\end{inparaenum}
%The PCC measures the linear relationship between the target business' ``check-ins'' and its neighbors' popularity indicator (\textit{i.e.}, either the total check-kins or the total likes).
We only use the number of ``check-ins'' for the target business %(and ignore the number of ``likes'')
because we are only interested in the physical presence of customers for the target business.
But for the target business's neighbors, we use both ``check-ins'' and ``likes'', as they
reflect the popularity---physical or metaphysical---of the area in which the target business is located.
% % <para connect>
% The PCC outputs a value between $+1$ and $-1$ inclusive, 
% where 1 is total positive correlation, 0 is no correlation, and $-1$ is total negative correlation.
For the neighbors' total ``check-ins'' and total ``likes'', 
we further partitioned them based on the relative distance from the target business.
Specifically, for every target business, we calculate the PCC between its ``check-ins'' 
and the total ``check-ins'' or ``likes'' of its neighbors within radius $r$,
where $r=\{50,100,150,200,\ldots,500\}$.

Figure~\ref{figure:pcc} shows the PCC of the two popularity indicators, broken down by 
the relative distance.
It is evident that, between the neighbors' ``check-ins'' and the neighbors' ``likes'',
the ``check-in'' feature is the better indicator as it has a higher PCC score than ``likes.''
Furthermore, nearer ``check-ins'' (\textit{e.g.}, 50 meters) have better PCC than further ``check-ins'',
which suggests that the nearer a target business is 
to a popular neighbor, the more ``check-ins'' it reaps.
On the contrary, the PCC score for ``likes'' increases as the distance between a target business and its surrounding neighbors increases.
This can be attributed to the nature of ``likes'', which reflects an online support for the business
and is not limited to physical proximity, 
whereas ``check-ins'' represent the registration of a person's physical presence,
which is determined by physical proximity.
% And although the ``likes'' count is less representative than the ``check-in'' count,
% the overall positive PCC scores suggest that regardless of ``likes'' or ``check-ins'', 
% in Singapore's food business, 
% there is still a ``local'' effect whereby businesses benefit from the presence of more
% popular, successful, and established neighbors.
% However, the same cannot be said for all business types; 
% for example, news businesses may find the ``like'' count more useful,
% though it is safe to say that businesses with physical stores will find ``check-in'' counts
% to be more representative.

%%%%%%%%%%%%%
%           %
% Framework %
%           %
%%%%%%%%%%%%%
\section{Proposed Framework}
\label{sec:framework}

Our location analytics framework, as illustrated in Figure ~\ref{figure:framework_overview}, 
consists of two phases:
\textit{training} and \textit{prediction}.
% <para connect>
%The training phase represents each business profile by a set of features.
%The set of profiles with ``check-ins'' are then used to learn the parameters of the prediction models.
%In this paper, we utilize gradient boosting and other regression methods.
%In the prediction phase, we analyze a target business profile
The training phase involves extracting a set of features from the existing business profiles and feeding them to a machine learning algorithm (\textit{i.e.}, gradient boosting \cite{Friedman:2001}; see Section \ref{sec:model}) in order to a predictive model for the ``check-in'' count.
In turn, the prediction phase involves extracting features from a target business profile 
and invoking the trained predictive model to generate a (predicted) ``check-in'' count for that profile. 
Note that the training phase is carried out \emph{offline}, whereas the prediction phase is done \emph{on the fly} for a (new) target profile.

The modules in our proposed framework consist of three main types:
\begin{inparaenum}[\upshape(\itshape i\upshape)]
\item \emph{input profiles},
\item \emph{feature extractor}, and
\item \emph{predictive model}.
\end{inparaenum}
We shall describe each module type in turn.

\subsection{Input Profile}

\begin{figure}[!t]
\centering
\includegraphics[width=1.0\columnwidth]{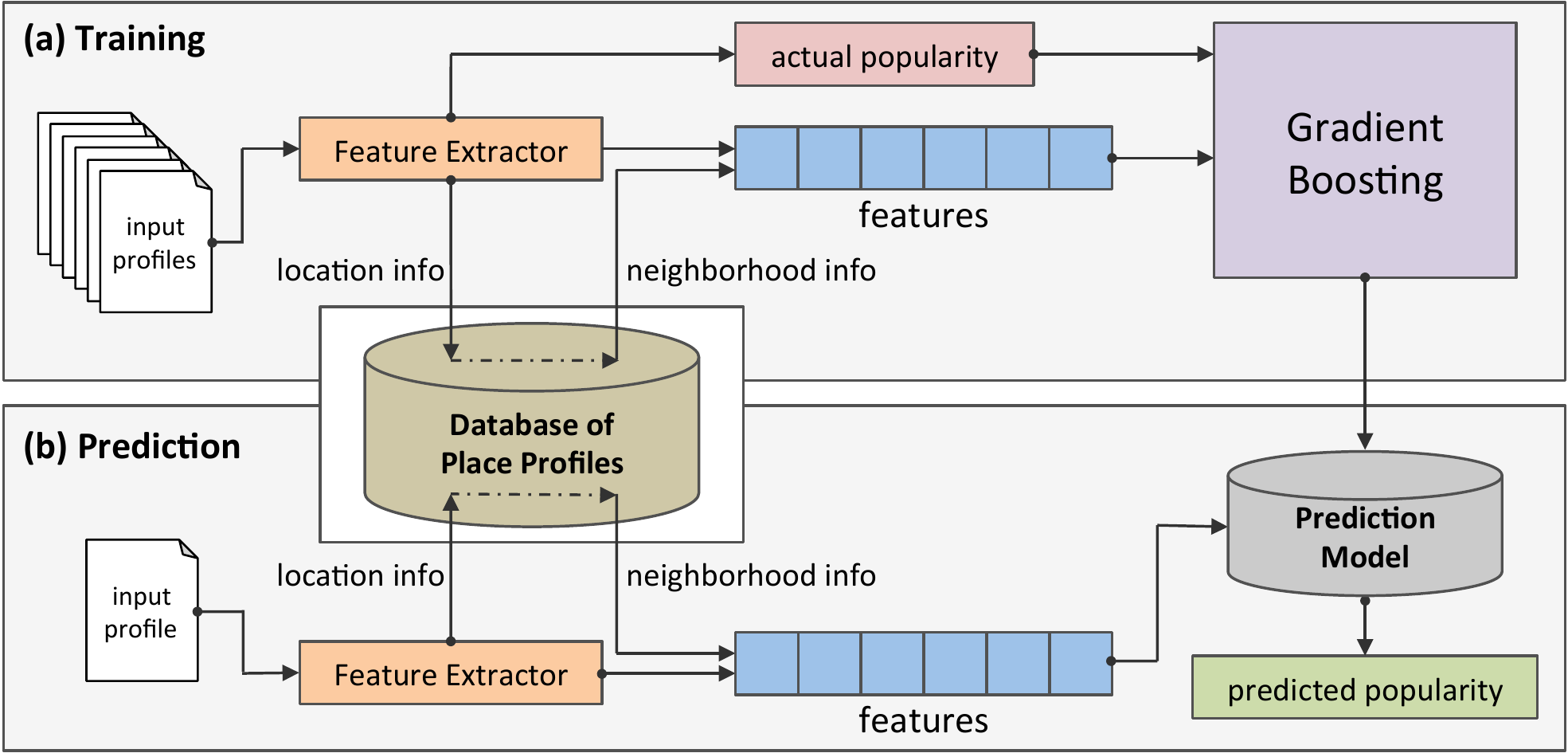}
\caption{Our location analytics framework.}
\label{figure:framework_overview}
\end{figure}

The input profile represents a physical business, and is used in both \textit{training} (Figure~\ref{figure:framework_overview}(a)) and \textit{prediction} (Figure~\ref{figure:framework_overview}(b)) phases.
An input profile contains several attributes of a business, namely:
% <para connect>
%\textbf{   (Ee-Peng: There is a contradiction. 
%            We just said that the profile is needed in both training and prediction phases.  
%            I suggest we drop ID and keep the rest which are actually needed in both training and prediction.)
%}
% <para connect>
\begin{itemize}
%\item The ID of the business profile ---
%this is only required in the \textit{training} phase as we are using
%existing profiles to train the prediction model;
%it is not needed in the \textit{prediction} phase as
%we are predicting the ``check-ins'' at a location of a hypothetical business.
\item \textbf{The lat-long coordinate of the business}. This is used in both training and prediction phases. Note that, during training, we only use the lat-long of the existing business profiles, whereas during prediction the lat-long being queried can be at an arbitrary (new or existing) location. %--- also needed in both training and prediction phases.

\item \textbf{The business categories}. Examples include ``bar'', ``cafe'', ``dining'', ``train station'', etc.. This information is also used in both training and prediction phases. % --- needed in both training and prediction phases.

\item \textbf{The ``check-in'' counts}. In the training phase, we feed the actual ``check-in'' counts of the existing businesses as the target variable of our algorithm. In the prediction phase, the ``check-in'' counts of the queried locations are assumed to be unknown and our algorithm is supposed to predict them, whether it is for an existing or a new location. %--- only required in the \textit{training} phase.
\end{itemize}
% <para connect>
%Once the input profile is constructed,
%it is then fed into the feature extractor (in the next section).

All input profiles of the existing businesses are stored in a database of place profiles (\emph{cf.} Figure \ref{figure:framework_overview}). Using this database, we can extract a set of features for a given business, which include features derived from its own (input) profile as well as features from its neighbors (computed based on a range of radius as described in Section \ref{sec:location_data}). The next section describes our feature extraction procedure. 

%++++++++++++++++%
% Figure (START) %
%++++++++++++++++%
%++++++++++++++%
% Figure (END) %
%++++++++++++++%

%++++++++++++++++%
% Figure (START) %
%++++++++++++++++%
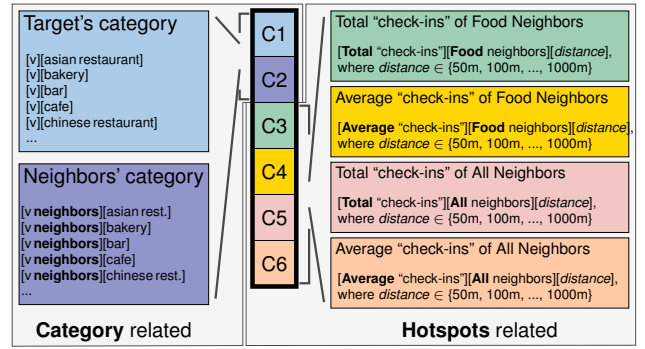
\begin{figure}[!t]
\pgfdeclarelayer{bg} 
\pgfsetlayers{bg,main}

% https://devimages.apple.com.edgekey.net/watch/human-interface-guidelines/specifications/images/design-fundamentals-color-palette.png
%\definecolor{color1}{HTML}{D6CEE1}
%\definecolor{color2}{HTML}{A0C3E7}
%\definecolor{color3}{HTML}{CEDBAD}
%\definecolor{color4}{HTML}{FBCBA4}
%\definecolor{color5}{HTML}{EBC6C5}
%\definecolor{color6}{HTML}{CFC8A9}
\definecolor{color1}{RGB}{172,203,232}
\definecolor{color2}{RGB}{146,149,202}
\definecolor{color3}{RGB}{158,206,180}
\definecolor{color4}{RGB}{255,214,2}
\definecolor{color5}{RGB}{242,198,197}
\definecolor{color6}{RGB}{255,202,165}

\definecolor{grey}{HTML}{F5F5F5}
\definecolor{grey2}{HTML}{505050}

\tikzstyle{macro_node}=[rectangle, draw=none, text centered, inner sep=2pt, text width=15pt, minimum height=20pt]
\tikzstyle{micro_node_cat}=[rectangle, draw=black, inner sep=2pt, text width=80pt, minimum height=57pt]
\tikzstyle{micro_node_hs}=[rectangle, draw=black, inner sep=2pt, text width=128pt, minimum height=28pt]
\tikzstyle{background}=[fill=grey,draw=grey2]

\tikzstyle{change_view_line}=[-,black, opacity=0.7, line width=1pt]

\adjustbox{max width=\columnwidth}{
\begin{tikzpicture}[font=\sffamily\sansmath]
	\node (chunk1-A) [macro_node, fill=color1, anchor=north] at (2.5,0) {C1};
	\node (chunk2-A) [macro_node, fill=color2, below=0pt of chunk1-A] {C2};
	\node (chunk3-A) [macro_node, fill=color3, below=0pt of chunk2-A] {C3};
	\node (chunk4-A) [macro_node, fill=color4, below=0pt of chunk3-A] {C4};
	\node (chunk5-A) [macro_node, fill=color5, below=0pt of chunk4-A] {C5};
	\node (chunk6-A) [macro_node, fill=color6, below=0pt of chunk5-A] {C6};
	\node (chunks-A-border) [draw=black, line width=2pt, inner sep=0pt, fit=(chunk1-A) (chunk2-A) (chunk3-A) (chunk4-A) (chunk5-A) (chunk6-A)] {};
	\draw [-,black] (chunk1-A.south west) -- (chunk1-A.south east);
	\draw [-,black] (chunk2-A.south west) -- (chunk2-A.south east);
	\draw [-,black] (chunk3-A.south west) -- (chunk3-A.south east);
	\draw [-,black] (chunk4-A.south west) -- (chunk4-A.south east);
	\draw [-,black] (chunk5-A.south west) -- (chunk5-A.south east);
	
	\node (chunk1-B) [micro_node_cat, fill=color1, anchor=north] at (0,0) {
\footnotesize Target's category
\fontsize{6}{7}\selectfont
\begin{adjustwidth}{-0.15cm}{}
\begin{tabular}{p{2.2cm} }
\\
{[}v{]}{[}asian\,restaurant{]}\\
{[}v{]}{[}bakery{]}\\
{[}v{]}{[}bar{]}\\
{[}v{]}{[}cafe{]}\\
{[}v{]}{[}chinese\,restaurant{]}\\
... 
\end{tabular}
\end{adjustwidth}
	};	
	\node (chunk2-B) [micro_node_cat, fill=color2, below=5pt of chunk1-B] {
\footnotesize Neighbors' category
\fontsize{6}{7}\selectfont
\begin{adjustwidth}{-0.23cm}{}
\begin{tabular}{p{2.2cm} }
\\
{[}v\,\textbf{neighbors}{]}{[}asian\,rest.{]}\\
{[}v\,\textbf{neighbors}{]}{[}bakery{]}\\
{[}v\,\textbf{neighbors}{]}{[}bar{]}\\
{[}v\,\textbf{neighbors}{]}{[}cafe{]}\\
{[}v\,\textbf{neighbors}{]}{[}chinese\,rest.{]}\\
... 
\end{tabular}
\end{adjustwidth}
	};
	\node (cat-related) [draw=none, below=4.3pt of chunk2-B] {\textbf{Category} related};
	
	\draw [change_view_line] ($(chunk1-A.north west) - (0.05,0.05)$) -- ($(chunk1-A.north west) - (0.2,0.05)$) -- ($(chunk1-A.north west) - (0.2,0.35)$);
	\draw [change_view_line] ($(chunk2-A.south west) - (0.05,-0.05)$) -- ($(chunk2-A.south west) - (0.2,-0.05)$) -- ($(chunk2-A.south west) - (0.2,-0.2)$);
	\draw [change_view_line] ($(chunk1-A.south west) - (0.2,-0.2)$) -- ($(chunk1-B.north east) + (0.1,-0.1)$);
	\draw [change_view_line] ($(chunk2-A.north west) - (0.2,0.2)$) -- ($(chunk2-B.south east) + (0.1,0.1)$);
	
	\node (chunk3-B) [micro_node_hs, fill=color3, anchor=north] at (5.7,0) {
\scriptsize Total ``check-ins'' of Food Neighbors
\fontsize{6}{7}\selectfont
\begin{adjustwidth}{-0.15cm}{}
\begin{tabular}{p{5cm}}
{[}\textbf{Total} ``check-ins''{]}{[}\textbf{Food} neighbors{]}{[}\textit{distance}{]},\\
where \textit{distance} $\in$ \{50m, 100m, ..., 1000m\}
\end{tabular}
\end{adjustwidth}
	};
	\node (chunk4-B) [micro_node_hs, fill=color4, below=2pt of chunk3-B] {
\scriptsize Average ``check-ins'' of Food Neighbors
\fontsize{6}{7}\selectfont
\begin{adjustwidth}{-0.15cm}{}
\begin{tabular}{p{5cm}}
{[}\textbf{Average} ``check-ins''{]}{[}\textbf{Food} neighbors{]}{[}\textit{distance}{]},\\
where \textit{distance} $\in$ \{50m, 100m, ..., 1000m\}
\end{tabular}
\end{adjustwidth}
	};
	\node (chunk5-B) [micro_node_hs, fill=color5, below=2pt of chunk4-B] {
\scriptsize Total ``check-ins'' of All Neighbors
\fontsize{6}{7}\selectfont
\begin{adjustwidth}{-0.15cm}{}
\begin{tabular}{p{5cm}}
{[}\textbf{Total} ``check-ins''{]}{[}\textbf{All} neighbors{]}{[}\textit{distance}{]},\\
where \textit{distance} $\in$ \{50m, 100m, ..., 1000m\}
\end{tabular}
\end{adjustwidth}
	};
	\node (chunk6-B) [micro_node_hs, fill=color6, below=2pt of chunk5-B] {
\scriptsize Average ``check-ins'' of All Neighbors
\fontsize{6}{7}\selectfont
\begin{adjustwidth}{-0.15cm}{}
\begin{tabular}{p{5cm}}
{[}\textbf{Average} ``check-ins''{]}{[}\textbf{All} neighbors{]}{[}\textit{distance}{]},\\
where \textit{distance} $\in$ \{50m, 100m, ..., 1000m\}
\end{tabular}
\end{adjustwidth}
	};
	\node (checkins-related) [draw=none, below=3pt of chunk6-B] {\textbf{Hotspots} related};
	
	\draw [change_view_line] ($(chunk3-A.north east) + (0.05,-0.05)$) -- ($(chunk3-A.north east) + (0.2,-0.05)$) -- ($(chunk3-A.north east) + (0.2,-0.5)$);
	\draw [change_view_line] ($(chunk6-A.south east) + (0.05,0.05)$) -- ($(chunk6-A.south east) + (0.2,0.05)$) -- ($(chunk6-A.south east) + (0.2,0.9)$);
	\draw [change_view_line] ($(chunk4-A.south east) + (0.2,0.2)$) -- ($(chunk3-B.north west) - (0.1,0.1)$);
	\draw [change_view_line] ($(chunk5-A.north east) + (0.2,-0.2)$) -- ($(chunk6-B.south west) - (0.1,-0.1)$);
	
	\begin{pgfonlayer}{bg}    % select the background layer
		\filldraw[background] ($(chunk2-A.south west)-(0.15,0)$) -- ($(chunk2-A.south east) + (0.1,0)$) -- ($(chunk1-A.north east) + (0.1,0.1)$) -- ($(chunk1-B.north west) + (-0.1,0.1)$) 
		-- ($(chunk2-B.south west) - (0.1,0.642)$) -- ($(chunk6-A.south west) - (0.15,0.92)$) -- ($(chunk2-A.south west)-(0.15,0)$);
		\filldraw[background] ($(chunk3-A.north west)-(0.1,0)$) -- ($(chunk3-A.north east) + (0.15,0)$) -- ($(chunk1-A.north east) + (0.15,0.1)$) -- ($(chunk3-B.north east) + (0.1,0.1)$)
		-- ($(chunk6-B.south east) + (0.1,-0.572)$) -- ($(chunk6-A.south west) - (0.1,0.92)$) -- ($(chunk3-A.north west)-(0.1,0)$);
	\end{pgfonlayer}
\end{tikzpicture}
}
\caption{A break down of the feature vector into six chunks.}
\label{figure:feature_vector}
\end{figure}
%++++++++++++++%
% Figure (END) %
%++++++++++++++%

\subsection{Feature Extraction}
\label{subsection:feature_extraction}

The feature extraction module serves to construct a feature vector representing a particular business. 
In this work, we divide our feature vector into six \emph{chunks}, which represent different aspects of a target business.
Figure~\ref{figure:feature_vector} summarizes our feature chunks. The first two are associated with \textit{categorical} data,
while the remaining four are about \textit{hotspots} (\textit{i.e.}, location and ``check-ins'') data.
Table \ref{table:feature_chunks} summarizes the unique identifier (ID), description, and the number of features of each chunk. We describe each chunk below.

%The feature extraction module in our framework can be further broken down
%into two sub-modules:
%\textit{categories} and \textit{neighborhood} extractions.

%\textbf{Categories Extraction}.
%\textit{Category}-related features can be separated into:
%\begin{inparaenum}[\upshape(\itshape i\upshape)]
%\item the \textit{target business}'s categories and
%\item its \textit{neighbors} categories.
%\end{inparaenum}

%\textbf{Neighborhood Extraction}.
%In contrast, \textit{hotspot}-related features can be separated into:
%\begin{inparaenum}[\upshape(\itshape i\upshape)]
%\item the \textit{total} ``check-ins'' of \textit{food}-only neighbors,
%\item the \textit{average} ``check-ins'' of \textit{food}-only neighbors,
%\item the \textit{total} ``check-ins'' of \textit{all} neighbors (\textit{i.e.}, food and non-food), %and
%\item the \textit{average} ``check-ins'' of \textit{all} neighbors.
%\end{inparaenum}

%Figure~\ref{figure:feature_vector} shows the feature vector of a business profile,
%which is made up of a synthesized subset of its attributes from Facebook's Graph API.
%The vector representation can be identified by 6 distinct chunks:

\textbf{Chunk $C_1$: The categories of the target business}.
This chunk is represented using a binary feature vector.
For example, a categorical variable with four possible values:
``A'', ``B'', ``C'', and ``D''
is encoded using four binary features:
$[1,0,0,0]$, $[0,1,0,0]$, $[0,0,1,0]$, and $[0,0,0,1]$, respectively.
To represent multiple categories, we simply use ``0'' and ``1'' to indicate the 
absence and presence of each category label respectively.
For example, we represent a profile with categories ``A,~C'' and another with categories ``A,~B,~C,~D'' as $[1,0,1,0]$ and $[1,1,1,1]$, respectively.
% Jovian added the following:
In other words, we use a one-vs-all scheme where we convert multi-class labels to binary labels
(\textit{i.e.}, belong or does not belong to the class).
As there are a total of 357 unique categories in the dataset of food venues,
the binary feature vector will have 357 elements.
%\richard{[Jovian: This is quite strange. In our web application, we only specify the lat-long (i.e., drop a pin) but never seem to state the categories. How can we compute chunk $C_1$ for a new unseen location?]}
% Jovian, we *are* able to give category information. It's at the top-right corner of the prototype.

\textbf{Chunk $C_2$: The categories of the target business' neighbors}. 
We first select---from our database of place profiles---the neighboring food businesses within $r$ meters from the target business, after which we extract and sum up the category feature vectors of the neighbors. 
To define category neighborhood, we use $r = 200$ meters, which we found to give optimal performance in our experiments.
%\footnote{Unless otherwise specified, $r$ is set to 200 meters as this radius is estimated 
%to be the optimal size of a neighborhood \cite{mehaffy2010urban}.}
Similar to $C_1$, chunk $C_2$ is also a 357-long feature vector that corresponds to the same number of unique categories, except that each feature value is now an integer.
Returning to our toy example of the four categories ``A'', ``B'', ``C'', and ``D'', if a profile only has 5 neighbors of category ``A'' and 7 neighbors of category ``B'', its integer feature vector will be $[5,7,0,0]$.

\textbf{Chunks $C_3$ and $C_4$: Food-related hotspots}.
The two chunks are related in that both only use \textit{food-related} neighbors.
In other words, they exclude neighbors that have no relevance to food,
such as clothing and electronic stores.
For each chunk, we are interested in ``hotspots'', which are circular areas with the profile in the center and each area is quantified by the ``popularity'' of stores within it. 
We define 20 hotspots around the profile whereby each hotspot is demarcated by a
maximum distance of $r$ meters, of which $r \in \{50,100,150,\ldots,1000\}$.
Finally, the only difference between $C_3$ and $C_4$ is in how ``popularity'' is defined;
the former computes the (natural) logarithm of the \textit{total} ``check-ins'' within a hotspot,
while the latter computes the logarithm of the \textit{average} ``check-ins''. 

It must be noted that the total and average ``check-ins'' include only the ``check-in'' counts of the neighbors and not the count of the target business itself (which is assumed to be unknown). 
Also, the purpose of applying logarithmic transformation to the ``check-in'' counts is to reduce the \emph{skewness} in the counts distribution (\emph{i.e.}, most businesses have small ``check-ins'' counts, but there is a handful number of businesses with unusually large ``check-ins''). 
In other words, applying logarithm transformation would allow us to mitigate the impact of (unusually) high ``check-ins'' for popular businesses.
%  we care is applied to capture both the tiny and large,
% with high acuity (high resolution) in sensing the tiny changes and
% coarse generalizations (low resolution) of large changes~\cite{buzsaki2014log}.
% That is, a logarithmic scale of ``check-ins'' would allow us to discern the ``check-in'' of $1$ than discern between ``check-ins'' of $100$ and $101$.
% https://www.quora.com/How-are-logarithms-used-in-real-life
% http://www.nature.com/nrn/journal/v15/n4/full/nrn3687.html
%In addition, it is easy to convert the popularity score back to the number of ``check-ins''---we simply calculate the exponential of the popularity score.
Previously, we conducted an experiment that used the \textit{raw} ``check-ins'' instead of the logarithmic values. Indeed, we observed that using the logarithm values yielded lower prediction errors than using the raw counts. 
As such, we shall focus on the results of the logarithm-scaled ``check-ins'' throughout the rest of this paper.

\textbf{Chunks $C_5$ and $C_6$: All (food + non-food) hotspots}.
These chunks are similar to $C_3$ and $C_4$, respectively.
The only difference is that, instead of solely using \textit{food-related} neighbors,
chunks $C_5$ and $C_6$ use food \textit{and} non-food neighbors together.
The non-food neighbors include bookshops, transportation facilities like bus and train stations, furniture stores, universities, etc.
We include non-food hotspots so as to capture the \emph{complementary} (non-food) businesses within the neighborhood of a target business.
%\richard{[Jovian: Why not just use non-food hotspots, instead of (food + non-food) ones? One might ask if the latter induces ``overlapping'']}
% Jovian added the following:
%We include (food + non-food) hotspots instead of pure (non-food) hotspots here so that we can analyze the importance of each Chunk $C$ later through 
%ablation testing.

%++++++++++++++++%
% Table (START) %
%++++++++++++++++%
\begin{table}[!t]
\caption{Feature chunks used in our location analytics work.}
\label{table:feature_chunks}
	\centering
	\small
	\begin{tabular}{clc}
	\toprule
	\textbf{Chunk ID} & \textbf{Chunk Description} & \#\textbf{Features}\\
	\midrule
	$C_1$ & Categories of the profile & 357\\
  $C_2$ & Categories of the profile's neighbors & 357\\
	$C_3$ & Total ``check-ins'' of food-related hotspots & 20\\
  $C_4$ & Average ``check-ins'' of food-related hotspots & 20\\
	$C_5$ & Total ``check-ins'' of all hotspots & 20\\
	$C_6$ & Average ``check-ins'' of all hotspots & 20\\
	\bottomrule
	\end{tabular}
\end{table}
%++++++++++++++++%
% TAble (END) %
%++++++++++++++++%

\subsection{Predictive Model}
\label{sec:model}

In order to learn the association between the extracted features and ``check-in'' scores of a given business. we train a supervised regression model
called gradient boosting machine (GBM) \cite{Friedman:2001}. GBM is a machine learning algorithm that iteratively
constructs an ensemble of weak decision tree learners through \emph{boosting} mechanism. 
Specifically, the boosting procedure consists of training weak learners and adding them into a final strong model in a forward stage-wise manner.
By combining many weak learners that have high bias (\emph{i.e.}, high prediction error), GBM yields an accurate and robust predictive model that has a lower bias than its constituent weak learners  \cite{natekin2013gradient}.
The GBM allows for the optimization of arbitrary differentiable loss functions for classification and/or regression task. For the purpose of ``check-in'' regression, however, we shall focus on the \emph{least square} loss function \cite{Friedman:2001} in this work.

Another major benefit of using GBM is that it can automatically derive the so-called \emph{feature importance} metric \cite{Friedman:2001,natekin2013gradient}.
This provides an important mechanism to interpret the trained model and identify the key features that contribute substantially to the prediction of the target variable (\emph{i.e.}, ``check-in'' score).
In particular, each decision tree in the GBM intrinsically performs feature selection by choosing the appropriate split points.
This information can then be used to measure the importance of each feature. 
That is, the more often a feature is used in the split points of a tree, the more important that feature is. 
This notion can be extended to the tree ensemble by averaging the feature importance of each tree.
We further elaborate our feature importance analysis in Section \ref{subsection:rq3}.

% as features do not often contribute equally to predict the target response (\textit{i.e.}, ``check-in'' count).
% Thus, when interpreting a model, the first question is:
% \textit{what are those important features and how do they contribute in predicting the target response?}
% A GBM model derives the ``feature importance'' from the fitted regression trees,
% each intrinsically performing feature selection by choosing appropriate split points.

%%%%%%%%%%%%%%%%%%%%%%%%%%%%
%                          %
% Evaluation Preliminaries %
%                          %
%%%%%%%%%%%%%%%%%%%%%%%%%%%%
\section{Evaluation Preliminaries}
\label{sec:preliminaries}

We preface our evaluation proper by detailing the evaluation metrics and procedure, the baseline models against which we compare GBM, as well as the model variations we considered in our study.

\subsection{Evaluation Metrics and Procedure}
To measure how accurate our predicted ``check-in'' scores differ from the
actual (observed) scores, we use two popular regression quality metrics:
\emph{mean-squared logarithmic error} (MSLE) and
%root-mean-squared logarithmic error~(RMSLE), and
\emph{mean absolute logarithmic error} (MALE).
The MSLE and MALE metrics are respectively defined as:
\begin{align}
\label{eqn:msle}
MSLE &= \frac{1}{n} \sum^{n}_{i=1}  \left( \log(p_i + 1) - \log(a_i + 1) \right)^2 \\ 
\label{eqn:male}
MALE &= \frac{1}{n} \sum^{n}_{i=1} \left| \log(p_i + 1) - \log(a_i + 1) \right|
\end{align}
where $n$ is the number of samples in the test set, $p$ is the predicted ``check-ins'', and $a$ is the actual ``check-ins''.
The MSLE metric measures the averaged squared errors, which gives a higher penalty to large logarithmic differences $|\log(p_i + 1) - \log(a_i + 1)|$.
On the other hand, the MALE metric measures the averaged absolute errors, whereby
all the individual differences are weighted equally.

To assess the performance of our predictive model, we perform a 10-fold cross-validation procedure
whereby the dataset is randomly partitioned into 10 equal sized subsamples.
%Of the 10 subsamples,
A single subsample is retained as the validation data for testing our models,
while the remaining~9 subsamples are used as training data.
The cross-validation process is then repeated 10 times,
with each of the 10 subsamples used exactly once as the validation data.
We then report the averaged performance.

Finally, to test for the statistical significance of our results, we utilize the \emph{independent two-sample t-test} \cite{Press1988}. In particular, we look at the $p$-value of the t-test involving two performance vectors, at a significance level of $0.01$. If the $p$-value is less than $0.01$, we can conclude the performance difference is statistically significant. %(and vice versa).

\subsection{Baselines}

We compare GBM with several regression baseline algorithms.
To foster reproducibility of this work, our implementations of all these algorithms (including GBM) are
based on the \textit{scikit-learn} library \cite{Scikit2015}.
The following baselines are used in this work:

\begin{itemize}
\item \textbf{Distance-based nearest neighbors (DNN)}. This is a simple baseline that takes the logarithm of the average ``check-ins'' of the neighbors that reside within some radius $r$ of a target business location. DNN works based on a simple intuition: ``\emph{the more popular the neighborhood, the more popular the target location is going to be, all else being equal}''. We test on $r \in \{50,100,\ldots,500\}$ and found that DNN with $r = 100$m brings about the best results.

\item \textbf{Support vector regression with linear kernel (SVR-Linear)} \cite{Fan2008}. This method produces a linear regression model that depends only on a subset of the training data, since the cost function for building the model ignores any data points close to the model prediction. For this method, we set the cost parameter to $C=1$ and the epsilon parameter (for controlling epsilon-insensitive loss) to $\epsilon=0.1$, which give the best performance in our experiments.

\item \textbf{Support vector regression with radial basis function kernel (SVR-RBF)} \cite{Smola2004}. This is the same as SVR-Linear, except that now it uses a radial basis function (Gaussian) kernel. As with SVR-Linear, we use $C=1$ and $\epsilon=0.1$, which again constitute the optimal configuration for SVR-RBF.
\end{itemize}

% Among the three, DNN is considered the simplest algorithm
% as it simply takes the log of the average ``check-ins'' of the neighbors
% that lie in a disk of radius~$r$ meters around a target profile.
% It does not require the feature vectors mentioned in
% Section~\ref{subsection:feature_extraction}.
% We train these models on the actual popularity score (derived from actual ``check-ins'') and test them on cross-validation data.
% We configure the models as follows:

% \textbf{Distance-based nearest neighbors (DNN)}.
% %Considered the simplest algorithm in this work,
% The only parameter for DNN is $r$, which is the
% radius (in meters) around a target profile.

% \textbf{SVR-RBF and SVR-Linear} \cite{Smola2004,Fan2008}.
% For both SVR-RBF and SVR-Linear,
% we use the default cost and epsilon parameters $C=1$ and $\epsilon=0.1$.
% Note that SVR-RBF uses the RBF kernel while SVR-Linear uses, as the name suggests, the linear kernel.

Last but not least, we configure our GBM algorithm using the ``least squares'' loss function, a learning rate of 0.1, a maximum tree depth of 10, and
a maximum tree width of ``sqrt'' (\textit{i.e.}, the square root of the total number of features).
For the number of boosting iterations $N$,
we perform an exhaustive \textit{grid search} on $N \in \{100,200,\ldots,5000\}$
%through a manually specified subset of the hyperparameter space of a learning algorithm.
%and report the results which yields the best performance on the various evaluation metrics.
and found that $N=1000$ produces the best results.
Note that, in each boosting iteration, a new tree is created and added into the ensemble. As such, the number of boosting iterations $N$ is equal to the number of trees constructed.

\subsection{Model Variations}

To evaluate the contributions of different feature chunks, we construct a variation of the predictive models by enumerating all possible combinations of the six chunks (see Section~\ref{subsection:feature_extraction}).
That is, we construct all possible $2^6 - 1 = 63$ chunk combinations and build a predictive model for each combination. 
We represent a model variant using a binary array of length six, where chunk $C_i$ maps to the $i^{th}$ element in the array.
We use the notation ``$\text{[model\_name]}_{\text{xxxxxx}}$'' to represent a particular model variant, where $\text{x} \in \{0,1\}$.
For example, a GBM model using $C_1$, $C_2$, and $C_4$ is denoted as $\text{GBM}_{\text{110100}}$.
Note that DNN does not use this notation, since it works based on spatial distance only, instead of feature chunks. 
For SVR-Linear, SVR-RBF and GBM, we run experiments on all 63 variants and report the best results for each of the three methods.

%%%%%%%%%%%
%         %
% RESULTS %
%         %
%%%%%%%%%%%

\section{Results and Analysis}
\label{sec:results}

We now present our main experimental results.
Our experiments seek to answer several key research questions (RQs):
\begin{description}
  \item[\textbf{RQ1}:] How well can our predictive model (GBM) estimate the popularity (\emph{i.e.}, ``check-in'' scores) of business locations? 
  \item[\textbf{RQ2}:] What are the contributions of different feature chunks? How robust is our model against different feature combinations?
  \item[\textbf{RQ3}:] Do the important features found by our model make sense? What can we learn/conclude from them?
\end{description}

%%%%%%%
% RQ1 %
%%%%%%%
\subsection{Performance Assessment (RQ1)}

Table~\ref{table:rq1} compares the cross-validation performances (\emph{i.e.,} averaged MALE and MSLE) of different regression methods.
For SVR-Linear and SVR-RBF, we show both the ``full variant'' (\textit{i.e.}, $\text{SVR-Linear}_{\text{111111}}$ and $\text{SVR-RBF}_{\text{111111}}$)
as well as the variants that give the best results for the same method (\textit{i.e.}, $\text{SVR-Linear}_{\text{111000}}$ and $\text{SVR-RBF}_{\text{100011}}$).
We observe that GBM consistently and significantly outperforms other models (at $p < 0.01$), particularly against $\text{SVR-RBF}_{\text{100011}}$, which is the best among all the baselines.

%=======================%
% Table for RQ1 (START) %
%=======================%
\begin{table}[!t]
\centering
\small
\caption{Performance comparisons of different models.}
\label{table:rq1}
  \begin{tabular}{ l c c c } % use 'Y' for first column
  \toprule
  \textbf{Model} & \textbf{Feature Chunks} & \textbf{MALE} & \textbf{MSLE}\\ % & \textbf{RMSLE} \\
  \midrule
         DNN\textsubscript{r=100m} & - & 1.99305 & 7.27499\\ % & 2.69722 \\
  \midrule
        SVR-Linear\textsubscript{111000} & $\{C_1, C_2\}$ & 1.59072 & 4.25301\\ % & \textbf{2.06228}\\
        SVR-Linear\textsubscript{111111} & $\{C_1, C_2, C_3, C_4, C_5, C_6\}$ & 2.12345 & 7.35446\\ % & 2.71191\\ 
  \midrule
        SVR-RBF\textsubscript{100011} & $\{C_1, C_5, C_6\}$ & 1.47518 & 3.61863\\ % & \textbf{1.90227}\\
        SVR-RBF\textsubscript{111111} & $\{C_1, C_2, C_3, C_4, C_5, C_6\}$ & 1.53067 & 3.92219\\ % & 1.98045\\
  \midrule
         \textbf{GBM\textsubscript{111111}} & $\{C_1, C_2, C_3, C_4, C_5, C_6\}$ & \textbf{1.16362*} & \textbf{2.56924*}\\ % & \textbf{1.60288**} \\
    \bottomrule
    \multicolumn{4}{l}{*: significant at 0.01 with respect to SVR-RBF\textsubscript{100011}}
    \end{tabular}
%    
% \caption*{
%     \begin{tabular}{| l l |}
%     \hline
%     \footnotesize{\textbf{\underline{LEGEND}}} & \\
%     \footnotesize $C_1$: \underline{categories} of profile & \footnotesize $C_2$: \underline{categories} of profile's neighbors \\
%     \footnotesize $C_3$: \textit{total} ``check-ins'' of \underline{food-related} hotspots & \footnotesize $C_4$: \textit{average} ``check-ins'' of \underline{food-related} hotspots \\
%     \footnotesize $C_5$: \textit{total} ``check-ins'' of \underline{all} hotspots & \footnotesize $C_6$: \textit{average} ``check-ins'' of \underline{all} hotspots\\
%     \hline
%     \end{tabular}
% }
\end{table}
%=====================%
% Table for RQ1 (END) %
%=====================%

We can explain the results in terms of model complexity.
For instance, we can expect the simplest nearest neighbor method (\textit{i.e.}, DNN) to be beaten by other methods, as it only uses spatial distance.
We can also anticipate that SVR-RBF would outperform SVR-Linear, as the RBF kernel maps the original features into a high-dimensional space. 
This expanded feature space provides SVR-RBF with a greater representation power to model a much more complex relationship than SVR-Linear.
Finally, as GBM combines weak learners into a strong learner, the aggregate prediction of the ensemble is more accurate than the prediction of any of its constituent learners. 
This aggregation also provides GBM with more robustness to data overfitting, as compared to SVM-RBF.

Additionally, the results in Table~\ref{table:rq1} suggest that the two SVR methods are more sensitive to the variation of feature chunks, particularly to the presence of less relevant (or irrelevant) features. 
This can be attributed to the fact that each tree in GBM intrinsically performs feature selection, for which less important features are unlikely to be chosen and used in the ensemble.
Indeed, we can see that SVR-Linear with all six chunks (\emph{i.e.}, $\text{SVR-Linear}_{\text{111111}}$) is outperformed by the simpler SVR-Linear variant (\emph{i.e.}, $\text{SVR-Linear}_{\text{111000}}$) that uses only three chunks. Surprisingly, the former is also outperformed by even the DNN method. The same conclusion can be made by comparing $\text{SVR-RBF}_{\text{111111}}$ and $\text{SVR-RBF}_{\text{100011}}$.
On the contrary, GBM is more robust against inconsequential features. In fact, GBM generally improves its performance as we add more chunks, as we will see shortly in Section~\ref{subsection:rq2}.

%Finally, the average absolute difference (of $\text{GBM}_{\text{111111}}$)
%between predicted ``check-ins'' and actual ``check-in''
%is 87.23\footnote{This \textit{differs} from our ``popularity'' metric, which is the log of the ``check-ins''.}.
%The average number of ``check-ins'' in our 20,877 businesses is 2,182.93. 
%\textbf{Ee-Peng: This error seems to look large. reasonable?}

%%%%%%%
% RQ2 %
%%%%%%%
\subsection{Contribution of Feature Chunks (RQ2)}
\label{subsection:rq2}

The partitioning of the feature vectors into six chunks allows us to investigate
the contribution of each feature group.
Table~\ref{table:rq2}(a) lists the top $10$ GBM variants (out of $63$ possible variants), 
sorted in an ascending order of their MALE scores.
Similarly, Table~\ref{table:rq2}(b) lists the top 10 GBM variants, sorted in ascending order of their
MSLE scores.
Note that the top $10$ GBM variants happen to be the same for the two tables, except that they have slightly different ordering.
From the results, it is evident that $\text{GBM}_{\text{111111}}$ does not significantly outperform the other nine variants (at a significance level of $0.01$). 
This shows that GBM is robust against the variation of feature chunks. 
We can also see that the performance of the GBM improves as we add more feature chunks. Again, this can be attributed to the feature selection mechanism of each tree in the ensemble, which helps exclude irrelevant features.
%=======================%
% Table for RQ2 (START) %
%=======================%
\begin{table}[!t]
\centering
\small
\caption{Cross-validation results of the top 10 GBM variants.}
\label{table:rq2}
  \begin{subtable}{1.0\columnwidth}
  \centering
  \caption{MALE results of the top 10 GBM variants.}
  \label{tab:resulttable2a}
  \begin{tabular}{lccccccc} % use 'Y' for first column
  \toprule
                   & \multicolumn{6}{c}{\textbf{Feature Chunks}} & \\
  \cline{2-7}
    \textbf{Model} & $C_1$ & $C_2$ & $C_3$ & $C_4$ & $C_5$ & $C_6$ & \textbf{MALE}\\ 
  \midrule
    GBM\textsubscript{111111} & Yes & Yes & Yes & Yes & Yes & Yes & \textbf{1.163618}\\
    GBM\textsubscript{111100} & Yes & Yes & Yes & Yes & --  & --  & 1.172693\\
    GBM\textsubscript{111010} & Yes & Yes & Yes & --  & Yes & --  & 1.173910\\
    GBM\textsubscript{110011} & Yes & Yes & --  & --  & Yes & Yes & 1.175062\\
    GBM\textsubscript{101111} & Yes & --  & Yes & Yes & Yes & Yes & 1.177136\\
    GBM\textsubscript{101100} & Yes & --  & Yes & Yes & --  & --  & 1.182053\\
    GBM\textsubscript{100011} & Yes & --  & --  & --  & Yes & Yes & 1.184053\\
    GBM\textsubscript{111000} & Yes & Yes & Yes & --  & --  & --  & 1.184895\\
    GBM\textsubscript{110010} & Yes & Yes & --  & --  & Yes & --  & 1.189258\\
    GBM\textsubscript{101010} & Yes & --  & Yes & --  & Yes & --  & 1.191831\\
  \midrule
  \textbf{Count} & 10 & 6 & 7 & 4 & 7 & 4 & \\
  \bottomrule
  \end{tabular}
  \vspace{4mm}
  \end{subtable}
  \begin{subtable}{1.0\columnwidth} 
  \centering
    \caption{MSLE results of the top 10 GBM variants.}
    \label{tab:resulttable2b}
    \begin{tabular}{lccccccc} % use 'Y' for first column
      \toprule
                   & \multicolumn{6}{c}{\textbf{Feature Chunks}} & \\
      \cline{2-7}
       \textbf{Model}  & $C_1$ & $C_2$ & $C_3$ & $C_4$ & $C_5$ & $C_6$ & \textbf{MSLE}\\
         %& \multicolumn{1}{c}{\multirow{-2}{*}{\normalsize \textbf{MSLE}}} \\
         %& \multicolumn{1}{c}{\multirow{-2}{*}{\normalsize \textbf{RMSLE}}}\\
      \midrule
         GBM\textsubscript{111111} & Yes & Yes & Yes & Yes & Yes & Yes & \textbf{2.569236}\\ % & \textbf{1.602884}\\
         GBM\textsubscript{111010} & Yes & Yes & Yes & --  & Yes & --  & 2.608927\\ % & 1.615217\\
         GBM\textsubscript{101111} & Yes & --  & Yes & Yes & Yes & Yes & 2.609254\\ % & 1.615318\\
         GBM\textsubscript{111100} & Yes & Yes & Yes & Yes & --  & --  & 2.610255\\ % & 1.615628\\
         GBM\textsubscript{110011} & Yes & Yes & --  & --  & Yes & Yes & 2.615818\\ % & 1.617349\\
         GBM\textsubscript{101100} & Yes & --  & Yes & Yes & --  & --  & 2.627101\\ % & 1.620834\\
         GBM\textsubscript{100011} & Yes & --  & --  & --  & Yes & Yes & 2.628505\\ % & 1.621266\\
         GBM\textsubscript{111000} & Yes & Yes & Yes & --  & --  & --  & 2.653032\\ % & 1.628813\\
         GBM\textsubscript{110010} & Yes & Yes & --  & --  & Yes & --  & 2.660369\\ % & 1.631064\\
         GBM\textsubscript{101010} & Yes & --  & Yes & --  & Yes & --  & 2.667292\\ % & 1.633185\\
      \midrule
         \textbf{Count} & 10 & 6 & 7 & 4 & 7 & 4 & \\ 
    \bottomrule
    \end{tabular}
  \end{subtable}
\end{table}
%=====================%
% Table for RQ2 (END) %
%=====================%

Based on the binary representation of the six chunks, we can also calculate the relative significance of a chunk by counting the number of times
in which it is present (\textit{i.e.}, when the chunk is assigned the value of $1$).
The sum of each chunk's presence in the $10$ GBM variants is shown at the last row of Tables~\ref{table:rq2}(a) and \ref{table:rq2}(b), entitled ``Count''.
We see that the categories of the target business (\textit{i.e.}, chunk $C_1$) is present in all the
top 10 GBM variants, indicating that it is an essential feature.
This may seem to suggest that the nature of the business itself plays a pivotal role.
However, as described in Section~\ref{subsection:categories_data},
food businesses on Facebook may contain non-food labels such as ``airport'' and ``shopping mall''
(\textit{e.g.}, for a cafe located in the shopping mall of an airport).
In turn, this suggests that the ``environment'' around a selected business is also a key factor.
The method of chunk counting presented in this section is a coarse-grained analysis and is not sufficient to validate this conjecture. We will further analyze this in Section~\ref{subsection:rq3}, where we employ a more fine-grained analysis of the individual feature's importance.

Moving on, we also notice that the total ``check-ins'' chunks (\textit{i.e.}, chunks $C_3$ and $C_5$)
are ranked higher than the average ``check-ins'' (\textit{i.e.}, $C_4$ and $C_6$), \emph{i.e.}, the counts are 7/10 vs. 4/10.
This suggests that the total ``check-ins'' have more discriminatory power than the average ``check-ins'',
which could be due to the averaging failing to account for the number of business nearby.
On the other hand, total ``check-ins'' (of an area) gives a more accurate reflection of the \textit{potential human traffic}
that an area has.
Finally, we see no substantial performance difference between \textit{food-related} hotspots and \textit{all} (\textit{i.e.}, food + non-food) hotspots (both have a count sum of $7+4=11$). 
This implies that the presence of non-food-related categories does not contribute significantly to the prediction quality.

%%%%%%%
% RQ3 %
%%%%%%%
\subsection{Analysis of Feature Importance (RQ3)}
\label{subsection:rq3}

The analysis of the six \textit{chunks} of the feature vectors in the previous section represents a coarse-grained analysis.
To perform a more fine-grained analysis, we look into the full feature vector (with all six chunks included)
in the $\text{GBM}_{\text{111111}}$ model and try to compute the \textit{relative importance} of each individual feature.
GBM derives this automatically, by measuring how many times a feature is used in the split points of a tree~\cite{Friedman:2001} (see also Section \ref{sec:model}).
%Such feature importance analysis is crucial for model interpretation, as features are seldom equally relevant;
%often only a few have substantial influence on the response, while the vast majority are irrelevant.

\begin{figure*}
\centering
\begin{subfigure}{.49\textwidth}
  \centering
  \includegraphics[width=1.0\textwidth]{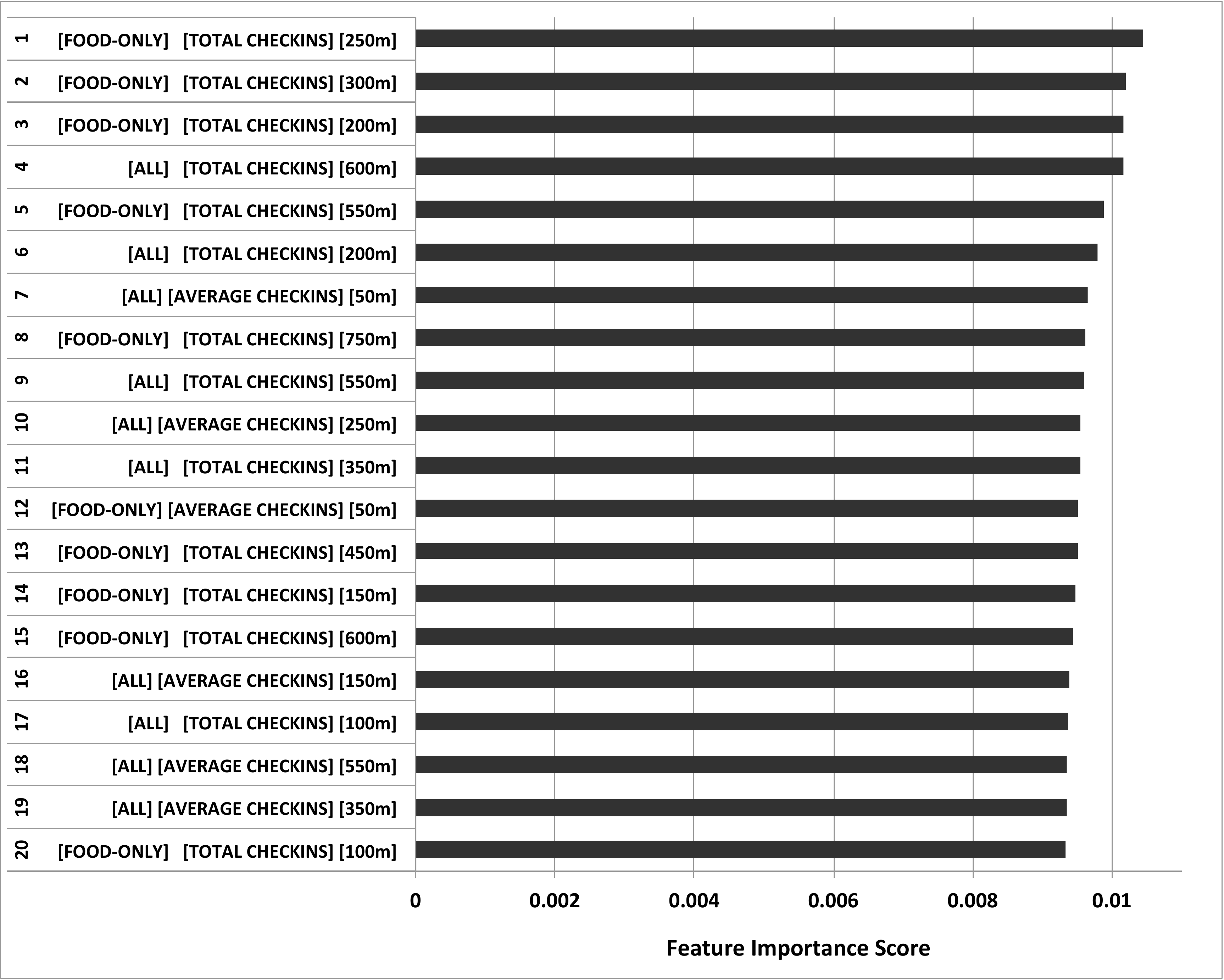}
  \caption{Feature importance of the top 20 features descending order \\of importance.
  The top 20 happen to consists of Chunks $C_3$ to $C_6$.}
  %\label{figure:XXX}
\end{subfigure}
\begin{subfigure}{.49\textwidth}
  \centering
  \includegraphics[width=1.0\textwidth]{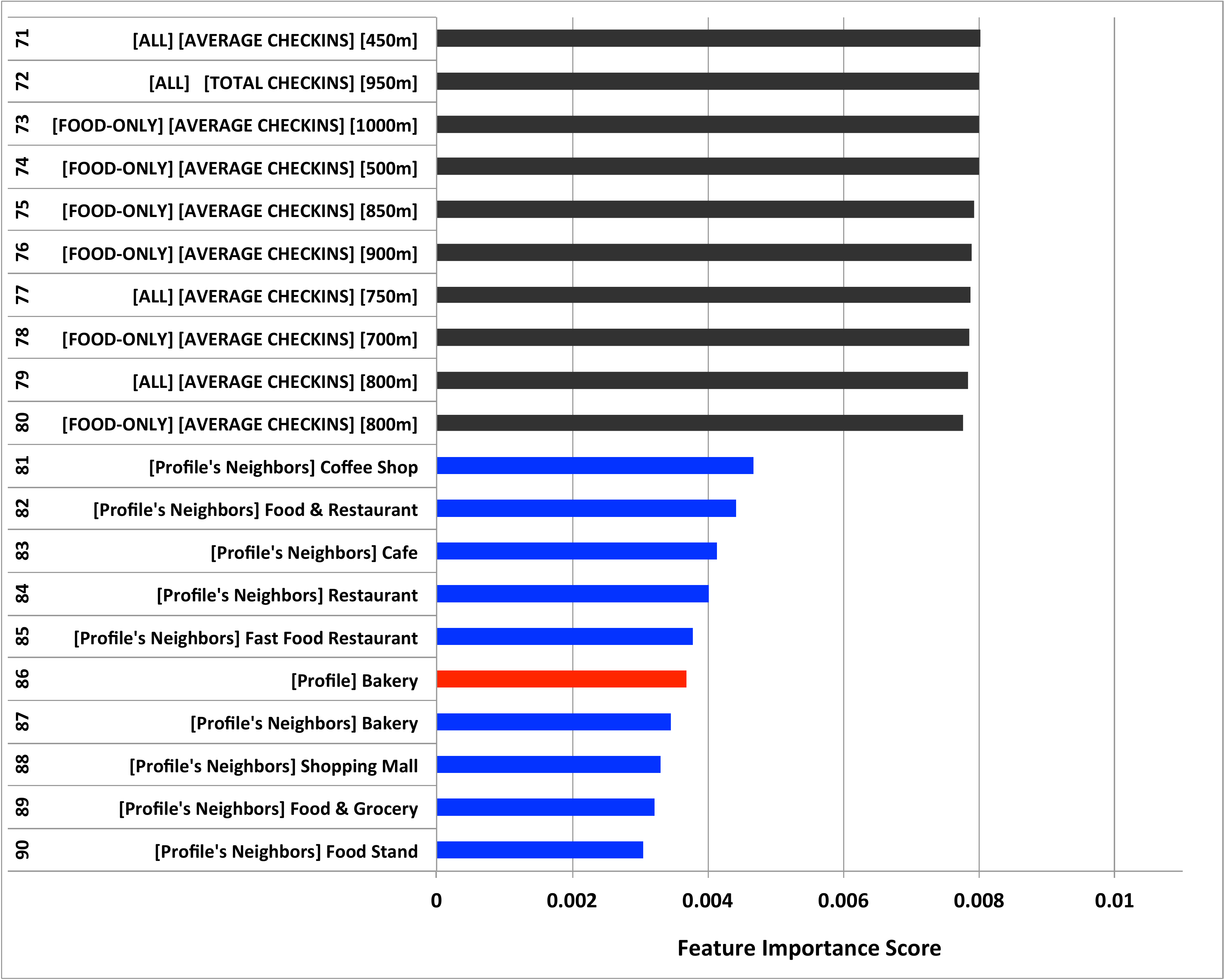}
  \caption{Feature importance of the top $\text{71}^{\text{st}}$ to $\text{90}^{\text{th}}$ features.
  Besides Chunks $C_3$ to $C_6$, it also contains Chunks $C_1$ and $C_2$, in red and blue, respectively.}
  %\label{figure:xxx}
\end{subfigure}
\caption{Top features from the feature importance of the $\text{GBM}_{\text{111111}}$ model.}
\label{figure:feature_importance}
\end{figure*}

Figure~\ref{figure:feature_importance}(a) shows the relative importance of the top 20 features in descending order of importance,
while Figure~\ref{figure:feature_importance}(b) shows the relative importance of the $71^{st}$ to the $90^{th}$
features. (We do not include the results for the $21^{st}$ to $70^{th}$ features here, as the changes in the feature importance score are fairly smooth.)
Accordingly, we can make the following observations:

\begin{itemize}
\item Chunks $C_3$ to $C_6$ (black bars in Figure~\ref{figure:feature_importance}) dominate the top 80 feature importance positions (not fully shown in Figure~\ref{figure:feature_importance}), and it is not until the $\text{81}^{\text{st}}$ top feature that chunks $C_1$ and $C_2$ show up.
This suggests that \emph{hotspot} features play a very crucial role: the more ``check-ins'' a target business' neighbors have,
the more popular the target business is likely to be. 

\item From Figure~\ref{figure:feature_importance}(a), 14 out of 20 hotspot features are below 500m, suggesting that nearer ``check-ins'' are used as a strong signal to make a split in the decision tree. This is not surprising, as it may be physically tiring for customers to travel farther than 500m, and most will settle all their outdoor needs in a specific area, such as a shopping mall.

\item Comparing the \textit{total} and \textit{average} ``check-ins'' in Figure~\ref{figure:feature_importance}(a),
14 out of 20 features belong to the former. This indicates that \textit{total} ``check-ins'' is a better input feature/signal
for split points in the GBM's trees. This finding is generally in agreement with what we have found in Section~\ref{subsection:rq2}.

\item Figure~\ref{figure:feature_importance}(a) also shows that the \textit{type} of neighbors (\textit{i.e.}, ``food-only'' or ``all'') are equally matched with 10 counts each. Again, this finding conforms with the earlier finding in Section~\ref{subsection:rq2}. Note that, despite the different approaches in Sections~\ref{subsection:rq2} and \ref{subsection:rq3}, both arrived at the similar conclusion with regard to the type of ``check-in'' and the type of neighbor.

\item From the colored bars (\textit{i.e.}, chunks $C_1$ and $C_2$) in Figure~\ref{figure:feature_importance}(b), we see that $C_1$ is dominated by the $C_2$. This suggests that  categories of the neighboring businesses are more important than those of the target business ($C_1$). Together with the ``hotspot'' features in Chunks $C_3$ to $C_6$, this reinforces the idea of a ``local effect'' whereby business benefit by being close to more established neighbors.

\item Finally, we notice a significant and faster drop in the \textit{importance} scores from the $\text{81}^{\text{st}}$ to  $\text{90}^{\text{th}}$ features (as compared from the $1^{st}$ to $80^{th}$). In this case, places or franchises that typically attract general (and larger) crowd, such as ``restaurant'', ``coffee'', or ``shopping mall''-related categories, take the first top spots among the neighbors' categories. This suggests that food-related categories (of the neighbors) are more important than the non-food categories.
\end{itemize}

%++++++++++++++++%
% Figure (START) %
%++++++++++++++++%
%++++++++++++++%
% Figure (END) %
%++++++++++++++%

% All in all, it can be con besides supporting our findings in Section~\ref{subsection:rq2}, 
% our \textit{feature importance} study sheds more light into our feature set, 
% providing a more detailed breakdown of the individual features and their significance.

%\section{Interactive Web Application}
\section{Web Application Prototype}
\label{sec:prototype}

%++++++++++++++++%
% Figure (START) %
%++++++++++++++++%
\begin{figure}[!t]
\centering
\setlength{\fboxsep}{0pt}
\setlength{\fboxrule}{0.5pt}
\fbox{\includegraphics[width=1\columnwidth]{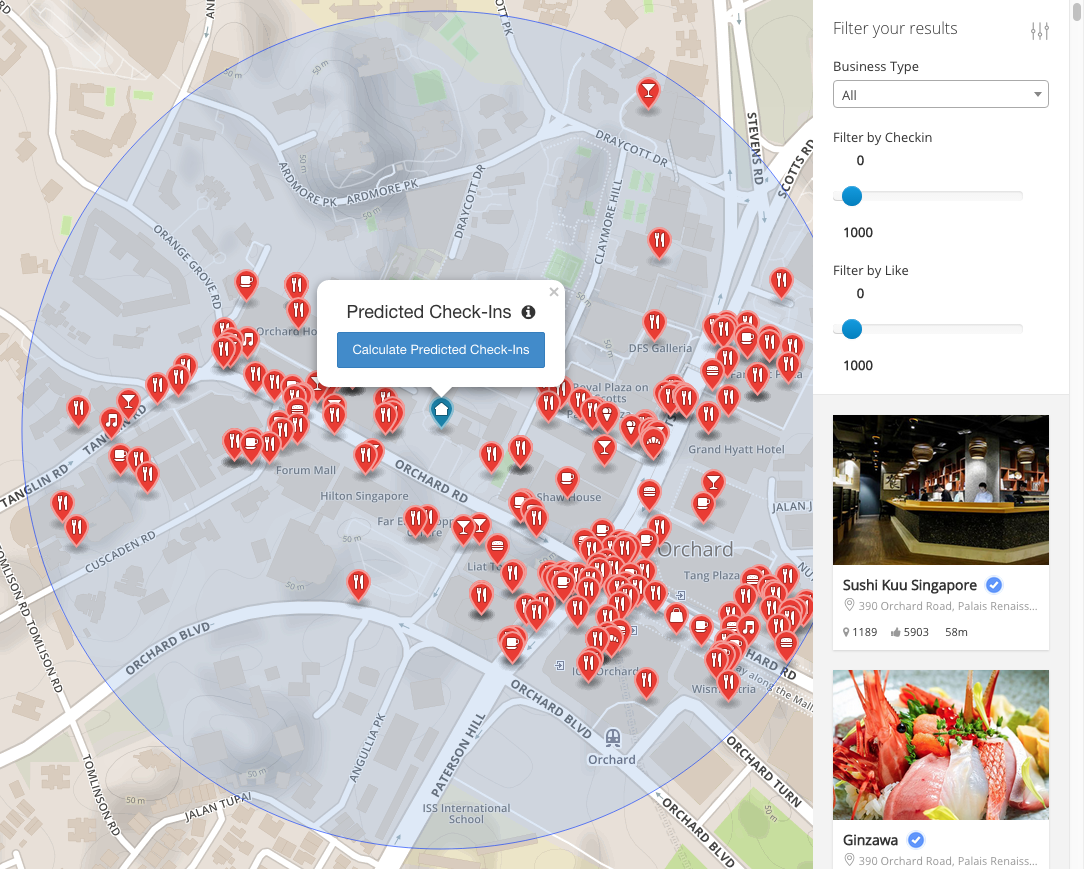}}
\caption{Our online interactive location analytics system.}
\label{figure:fb_analytics_prototype}
\end{figure}
%++++++++++++++%
% Figure (END) %
%++++++++++++++%
 
We implement our location analytics framework as an interactive web application service, which can be accessed at:
\url{https://research.larc.smu.edu.sg/bizanalytics/}. 

\textbf{Technologies}. We employ the following technologies to build our web application:
\begin{inparaenum}[\upshape(\itshape i\upshape)]
\item Python (implementing the predictive model and feature extraction),
\item RabbitMQ (a messaging passing system that allows querying the predictive model),  
\item Node.js (for processing users' queries and returning the prediction results to the front-end),
\item ElasticSearch (a distributed search engine for quering the database of place profiles), and
\item Google Maps (for visualization at the front-end).
\end{inparaenum}
This configuration provides an efficient and scalable way to process a user's location query (via Node.js and RabbitMQ), retrieve the relevant neighbors (via ElasticSearch), involve feature extraction and predictive models (in the Python component), and finally display the prediction results to the users (again via Node.js and RabbitMQ, along with Google Maps).
%\richard{[Jovian: Please also check if the above is correctly written.]}
% Jovian: OK

\begin{figure}[!t]
\centering
\setlength{\fboxsep}{0pt}
\setlength{\fboxrule}{0.5pt}
\fbox{\includegraphics[width=1\columnwidth]{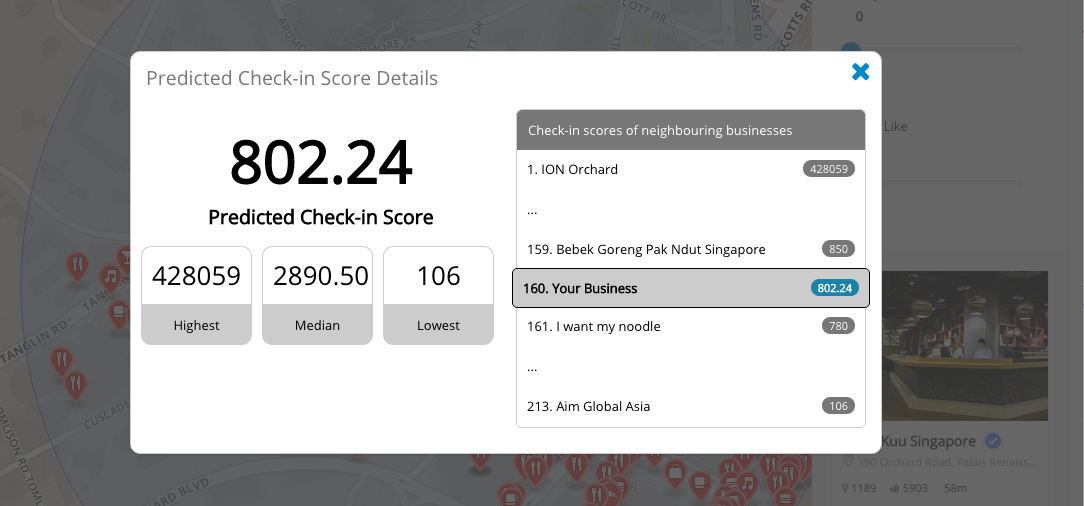}}
\caption{After computing the predicted ``check-in'' score, users can compare their target location with the surrounding businesses.}
\label{figure:fb_analytics_prototype_4}
\end{figure}

\textbf{User interaction}. Figure~\ref{figure:fb_analytics_prototype} shows an example of how our web application works. 
A user drops a pin (\emph{i.e.}, the blue pin in the middle of the screen) to indicate where his (hypothetical) store location would be. 
Depending on the location, the interface also dynamically shows the neighboring businesses on the right panel and their respective information, such as \emph{(i)} the distance from the drop-pin, and \emph{(ii)} the number of physical ``check-ins'' and ``likes''.
When the user is ready, he/she may click the ``Calculate Predicted Check-ins'' button, which will then calculate the predicted ``check-in'' score on the fly.
After presenting the predicted ``check-in'' score, the user can also open a new panel, showing a ranked list of his/her target location relative to the nearby businesses (Figure~\ref{figure:fb_analytics_prototype_4}).
The ranking allows users to understand how their target location would fare against the neighboring businesses.
The panel also shows the highest, lowest, and the median scores of these neighbors.
% In turn, our prediction model (\textit{i.e.}, GBM) will compute either a predicted ``check-in'' score, or the popularity score (\textit{i.e.}, log of the predicted ``check-ins''), as shown by the callout %above the pin location.

%++++++++++++++++%
% Figure (START) %
%++++++++++++++++%
%++++++++++++++%
% Figure (END) %
%++++++++++++++%

\textbf{Qualitative study}. 
Figures~\ref{figure:fb_analytics_prototype} and \ref{figure:fb_analytics_prototype_4}
demonstrate the on-the-fly prediction of our web application, where it is able to predict the ``check-in'' score of a
hypothetical, inexistent target business. 
For this example, the score of $802.24$ in Figure~\ref{figure:fb_analytics_prototype_4} represents a fairly conservative 
prediction of the potential ``check-ins'' in the target location (\emph{i.e.}, the blue pin) and the selected type of business.
This hypothetical business is ranked $160^{th}$ among $213$ businesses, with the lowest ``check-ins'' being $106$.
This is a reasonable prediction. On the one hand, because the place is near places with consistent human traffic, such as the Hilton Hotel and several other shopping malls, it should garner a decent amount of check-ins.
On the other hand, as there are many other businesses in the area (the area is a renowned shopping paradise in Singapore),
it may be challenging for the hypothetical business to compete with these businesses.

\section{Discussion and Future Work}
\label{sec:conclusion}

In this work, we investigate whether businesses can benefit from other (popular) businesses within its vicinity.
Our results show not only a positive correlation between the popularity of a target business and its neighbors, but also
the critical importance of the ``hotspot'' features: the nearer a target location is to a popular place with larger
``check-ins'', the more successful it would be. 
This finding conforms with our intuition. But more importantly, it demonstrates that ubiquitous online data (such as Facebook Pages) can be used to gauge the socioeconomical values. 
We also show how our predictive model can be used to accurately estimate the ``check-in'' score of a particular location, allowing us to identify the best locations that would bring popularity, and by extension, success.

Despite the promising potentials of our approach, there remains room for improvement. 
For instance, our current work has not taken into account the temporal aspects of the business popularity, such as modeling the trend of the ``check-in'' scores over time.
Further quantitative and qualitative studies may also be needed in the future to compare our work with other location-based services such as Foursquare.
To facilitate more comprehensive location analytics, we can extend our approach by building a two-level location recommendation system, whereby we first (coarsely) recommend a city district \cite{Lin2016} and then pinpoint (multiple) promising locations within that district. 
As we include more data, such as non-food categories and auxiliary data that reflect the human flow of different areas of an urban city, we will be able to further improve on our current model and findings.
To address all these, we plan to develop a new \emph{spatiotemporal} predictive model that integrates a richer set of residential, demographics, and other social media data.

\hfill

\noindent \textbf{Acknowledgements}. This research is supported by the National Research Foundation, Prime Minister's Office, Singapore under its International Research Centres in Singapore Funding Initiative.

%====================%
% *** References *** %
%====================%

\bibliographystyle{abbrv}
%\small
%\balance
\bibliography{main}
%\balancecolumns

% that's all folks
\end{document}